\documentclass[pre,superscriptaddress,showpacs,twocolumn,longbibliography]{revtex4-1}

\usepackage{hyperref}

\usepackage{color}
\usepackage[usenames,dvipsnames]{xcolor}
\usepackage{amsmath,amsthm,amssymb}
\usepackage{graphicx}
\usepackage{epsfig}
\usepackage{dcolumn}
\usepackage{bm}
\usepackage{mathrsfs}
\usepackage{multirow}
\usepackage[all]{xy}
\usepackage{pbox}
\usepackage{verbatim}

\DeclareMathOperator{\Tr}{Tr}
\def\(({\left(}
\def\)){\right)}
\def\[[{\left[}
\def\]]{\right]}

\newcommand{\be}{\begin{equation}}
\newcommand{\ee}{\end{equation}}
\newcommand{\ben}{\begin{eqnarray}}
\newcommand{\een}{\end{eqnarray}}
\newcommand{\beq}{\begin{equation}}
\newcommand{\eeq}{\end{equation}}

\begin{document}

\title{A non-equilibrium quantum many-body Rydberg atom engine}
\author{Federico Carollo}
\affiliation{School of Physics and Astronomy and \\Centre for the Mathematics and Theoretical Physics of Quantum Non-Equilibrium Systems, University of Nottingham, Nottingham, NG7 2RD, UK}
\author{Filippo M. Gambetta}
\affiliation{School of Physics and Astronomy and \\Centre for the Mathematics and Theoretical Physics of Quantum Non-Equilibrium Systems, University of Nottingham, Nottingham, NG7 2RD, UK}
\author{Kay Brandner}
\affiliation{Department of Physics, Keio University, 3-14-1 Hiyoshi, Yokohama, 223-8522, Japan}
\author{Juan P. Garrahan}
\affiliation{School of Physics and Astronomy and \\Centre for the Mathematics and Theoretical Physics of Quantum Non-Equilibrium Systems, University of Nottingham, Nottingham, NG7 2RD, UK}
\author{Igor Lesanovsky}
\affiliation{School of Physics and Astronomy and \\Centre for the Mathematics and Theoretical Physics of Quantum Non-Equilibrium Systems, University of Nottingham, Nottingham, NG7 2RD, UK}
\affiliation{Institut f\"ur Theoretische Physik, Universit\"at T\"ubingen, Auf der Morgenstelle 14, 72076 T\"ubingen, Germany}

\date{\today}

\begin{abstract}
The standard approach to quantum engines is based on equilibrium systems and on 
thermodynamic transformations between Gibbs states. 
However, non-equilibrium quantum systems offer enhanced experimental flexibility in the control of their parameters and, if used as engines, a more direct interpretation of the type of work they deliver. 
Here we introduce an out-of-equilibrium quantum engine inspired by recent experiments with cold atoms. Our system is connected to a single environment and produces mechanical work from many-body interparticle interactions arising between atoms in highly excited Rydberg states.
As such, it is not a heat engine but an isothermal one. 
We perform many-body simulations to show that this system can produce work. 
The setup we introduce and investigate represents a promising platform for devising new types of microscopic machines and for exploring  quantum effects in thermodynamic processes.
\end{abstract}

\maketitle 
Engines are devices able to convert some form of energy into mechanical work. The most famous examples, heat engines, operate by exchanging heat with (at least) two thermal reservoirs at different temperatures \cite{Martinez:2015aa,Krishnamurthy:2016aa,Rossnagel325,Josefsson:2018aa}; other working principles can be implemented also with a single reservoir \cite{Schliwa:2003aa,Seifert_2012,PhysRevLett.120.260601}. Nowadays, due to significant technological breakthroughs in manipulating and controlling microscopic systems, a new focus is on devising and realising efficient machines harnessing quantum effects \cite{PhysRevLett.120.260601,PhysRevLett.122.110601,PhysRevX.5.031044,PhysRevLett.112.030602,Jaramillo_2016,PhysRevB.99.024203}. To explore possible avenues at this scale, quantum thermodynamics has been put forward as a theoretical framework merging features of quantum physics with the laws of thermodynamics \cite{anders,Alicki2018,Deffner2019}. While much progress has been made in theoretically describing quantum engines, it is often not clear how energy, in the form of mechanical work, can be extracted from a many-body quantum system.

\begin{figure*}[t]
\centering
\includegraphics[scale=0.62]{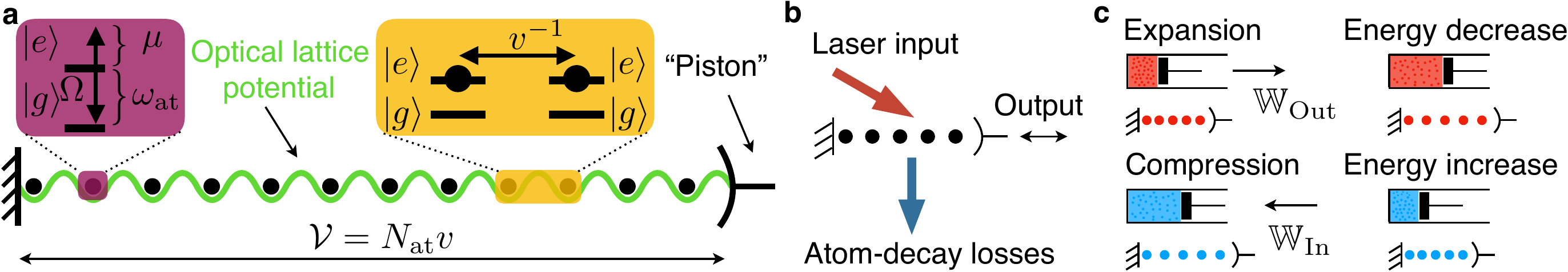}
\caption{{\bf Rydberg atomic chain and schematics of the engine.} {\bf a} Array of Rydberg atoms confined in a 1D geometry. The volume $\mathcal{V}$ is assumed to be tunable  through a movable ``piston" or mirror,  and the specific volume $v=\mathcal{V}/N_{\rm at}$ accounts for the interatomic distance. Each atom is modeled as a two-level system consisting of a ground state $|g\rangle$ and an excited (Rydberg) state $|e\rangle$. The term $\omega_{\rm at}$ is the single-atom energy difference between these two states. Laser driving generates oscillations, $|g\rangle \leftrightarrow |e\rangle$, at Rabi frequency $\Omega$; transitions can be made off-resonant through the detuning $\mu$. Two neighbouring excited atoms interact with a strength $v^{-1}$. 
{\bf b} Sketch of the Rydberg atom engine. The quantum system is driven out-of-equilibrium by a laser injecting energy which can be lost through atom decay. By means of an appropriate periodic manipulation of the ``piston", the system can deliver positive net output work. {\bf c} Four-stroke cycle and analogy with Rydberg atomic engine. During the expansion work $\mathbb{W}_{\rm Out}$ is extracted. This stroke is followed by a {\it cooling} step decreasing the interaction energy of the working fluid. Then a compression performs work $\mathbb{W}_{\rm In}$ on the system, and a {\it heating} step increases the interaction energy. Cooling and heating are implemented by varying the laser detuning.}
\label{Fig1}
\end{figure*} 

Here, we present and analyze a quantum engine whose working system is described by genuine non-equilibrium many-body steady-states and not by thermal states (see, e.g., \cite{PhysRevLett.108.085303,Chen:2019aa}), as is more standard. This novel feature comes at an additional cost in terms of efficiency:  
maintaining a non-equilibrium steady-state requires the constant injection of energy which is continuously dissipated. Current experiments allow for the implementation of  such driving protocol and for the precise control over the emerging non-equilibrium states. We present our ideas in the context laser driven Rydberg atoms [see Fig.~\ref{Fig1}(a)]
arranged in a one-dimensional (1D) chain, e.g. achieved by means of optical lattices or optical tweezer arrays \cite{Bloch:2005aa,PhysRevLett.107.263001,RevModPhys.80.885,RevModPhys.82.2313,Endres1024,Barredo1021,Labuhn:2016aa,PhysRevLett.118.063606,Bernien:2017aa,PhysRevX.8.021070,PhysRevX.8.041055}. In this scenario, we envisage for the sake of illustration a movable ``piston" subject to the (repulsive) force between Rydberg excited atoms and allowing one to tune the volume of the chain, as shown in Fig.~\ref{Fig1}. In practice this may indeed be realized as a cavity or rather an array of optical tweezers implemented with micromirror arrays \cite{Endres1024,Barredo1021,Labuhn:2016aa,PhysRevLett.118.063606,Bernien:2017aa,PhysRevX.8.021070,PhysRevX.8.041055}.
This physical setup offers  a transparent interpretation of the engine, sketched in Fig.~\ref{Fig1}(b), in particular in terms of the nature of the work it can provide. The laser pumps energy (input) into the system which converts it into interaction energy of Rydberg states, and which can then be extracted through the mechanical motion of the piston (output). The system thus acts as an opto-mechanical energy converter, 
with spontaneous decay of Rydberg excited states leading to constant energy losses during the cycle. The periodic protocol we consider consists of two isochoric transformations, which increase or decrease the density-density interactions of the Rydberg ``working fluid", and of two transformations where the volume is varied. This cycle is illustrated in Fig.~\ref{Fig1}(c) through an analogy with a classical  engine. 

Going beyond previous proposals, our engine is based on a genuine non-equilibrium protocol, which we  investigate from a fully dynamical, i.e.\ explicitly time-dependent, viewpoint \cite{PhysRevLett.108.085303,Chen:2019aa}.  Unlike recent work, see e.g.\ Refs.~\cite{PhysRevLett.108.085303,Niedenzu_2018,Li_2018}, we provide nonperturbative results for open many-body quantum systems  beyond the mean-field approximation. 
Moreover, our device is not  a heat engine but rather an isothermal engine \cite{Seifert_2012} operating far from equilibrium in contact with a single environment. From an experimental viewpoint this approach provides a substantial simplification since the engine does not need to alternate between two different heat baths \cite{Chen:2019aa}. Our setup can be used to design realistic quantum devices or as a new testbed to explore the impact of quantum effects on thermodynamic processes far from equilibrium. Finally, our framework provides, at least in principle, a viable mechanism for direct work measurement and extraction; the piston is a macroscopic object and all relevant quantities that characterise our engine are accessible experimentally by measuring  density correlation functions through spatially resolved imaging of Rydberg excitations \cite{Schaus:2012aa}.\\

\noindent {\bf \em The model-- } We consider a 1D array of laser driven Rydberg atoms. The Hamiltonian, in a frame rotating with the laser frequency, is given by \cite{PhysRevLett.108.023602} [{\it c.f.} Fig.~\ref{Fig1}(a)]
\begin{equation}
H=\Omega\sum_{k=1}^{N_{\rm at}} \sigma_{x}^{(k)}-\mu\sum_{k=1}^{N_{\rm at}} n^{(k)} +H^v_{\rm LG}\, .
\label{H}
\end{equation}
Here, $\sigma_x|g/e\rangle =|e/g\rangle$, while $n$ counts the presence of an excitation  $n|e\rangle =|e\rangle$ and $n|g\rangle =0$. ${N_{\rm at}}$ is the total number of atoms. The first two terms are related to the laser driving. The term $H_{\rm LG}^v$ is the {\it lattice gas} Hamiltonian accounting for classical (i.e.\ diagonal) volume-dependent (repulsive) nearest-neighbour interactions,
$$
H_{\rm LG}^v=\frac{1}{v}\sum_{ k=1 }^{N_{\rm at}-1}n^{(k)}n^{(k+1)}\, .
$$
This  term is directly related to the mechanical energy stored in the system which can be extracted or injected by varying the volume of the 1D array of atoms, as shown in Fig.~\ref{Fig1}(b-c).

The roles played by the other terms in the Hamiltonian \eqref{H} are as follows: the one proportional to $\Omega$ leads to excitations being created and annihilated. This parameter is not altered during the cycle; it rather provides the system with the background fluctuations needed to generate atomic transitions between $|g\rangle$ and $|e\rangle$. Transitions are further controlled by the detuning term $\mu$. Large detunings make transitions off-resonant, suppressing the probability amplitude of observing excitations or de-excitations. Combining this observation with the fact that the system also experiences  spontaneous atomic decays [see Eq.~\eqref{Lindblad} below], one would expect to observe, on average and after a transient, a larger number of Rydberg excitations for small detunings $\mu$. Indeed, when the detuning is large, atoms which are found in the ground state after decaying are less likely excited due to the transition being off-resonant. This effect leads to a lower Rydberg state population and, in turn, to a lower interaction energy. Hence, varying the parameter $\mu$ makes it possible to modify the interaction energy of the Rydberg system without changing its volume; this allows us to realise the engine cycle as discussed in Fig.~\ref{Fig1}(c).

In order to  account for the spontaneous decay of excited states, which is indeed a non-negligible feature of experiments, we exploit a Markovian dissipative map in Lindblad form \cite{lindblad76a,gorini1976}. The latter is defined for atom decay as 
\begin{equation}
\mathcal{L}[X]:=\gamma\sum_{k=1}^{N_{\rm at}}\left(\sigma^{(k)}_-X\sigma_+^{(k)}-\frac{1}{2}\left\{n^{(k)},X\right\}\right)\, .
\label{Lindblad}
\end{equation}
Here, $\gamma^{-1}$ represents the characteristic life-time of the Rydberg state, and $\sigma_-|e\rangle =|g\rangle$, $\sigma_+=\sigma_-^\dagger$. Altogether, the system  density matrix $\rho_t$ evolves, in the rotating frame, through the equation
\begin{equation}
\dot{\rho}_t=-i[H,\rho_t]+\mathcal{L}[\rho_t]\, .
\label{int-picture}
\end{equation}
This model for laser driven Rydberg systems is, in fact, phenomenological. Still, it is commonly used to interpret cold-atom experiments as it enables a quantitative description of atomic interactions and has been extensively tested in practice. Note that the dissipative generator of Eq.\eqref{Lindblad} contains local Lindblad operators even though the Hamiltonian $H$ involves interactions. This feature can lead to violations of the second law in generic instances \cite{Levy_2014,Barra:2015aa,PhysRevE.93.062134,Stockburger2017}. However, as we show below, in our {\it zero-temperature} setup, in which only atom decay plays a role \cite{PhysRevE.93.062134}, there are no thermodynamic inconsistencies. \\

\noindent {\bf \em Internal energy, first law and engine cycle--} 
In order to discuss the engine depicted in Fig.~\ref{Fig1} and the net output that it can deliver,  we need to develop a thermodynamic description of the system. To this end, we first have to identify the {\it internal energy} of the engine, which corresponds to the energetic content in the working fluid that could, in principle, be extracted in the form of work. In our setup, this quantity is related to the repulsive interaction energy $H^v_{\rm LG}$ and to the single-atom energy 
$$
H_{\rm at}=\omega_{\rm at} \sum_{k=1}^{N_{\rm at}}n^{(k)}\, ,
$$
associated with the energy difference, $\omega_{\rm at}$, between excited and ground state, see Fig.~\ref{Fig1}(a). We thus define the system energy operator, $H_{\rm in}^v$, as 
\begin{equation}
H_{\rm in}^v:=H_{\rm at}+H_{\rm LG}^{v}
\label{int-operator}
\end{equation}
and the specific internal energy as
\begin{equation}
E_{\rm in}(t):=\frac{1}{N_{\rm at}}\Tr\left[\rho_t \, H_{\rm in}^v\right]\big|_{v=v_t}\, .
\label{int-energy}
\end{equation}

\begin{figure}[t]
\centering
\includegraphics[scale=0.7]{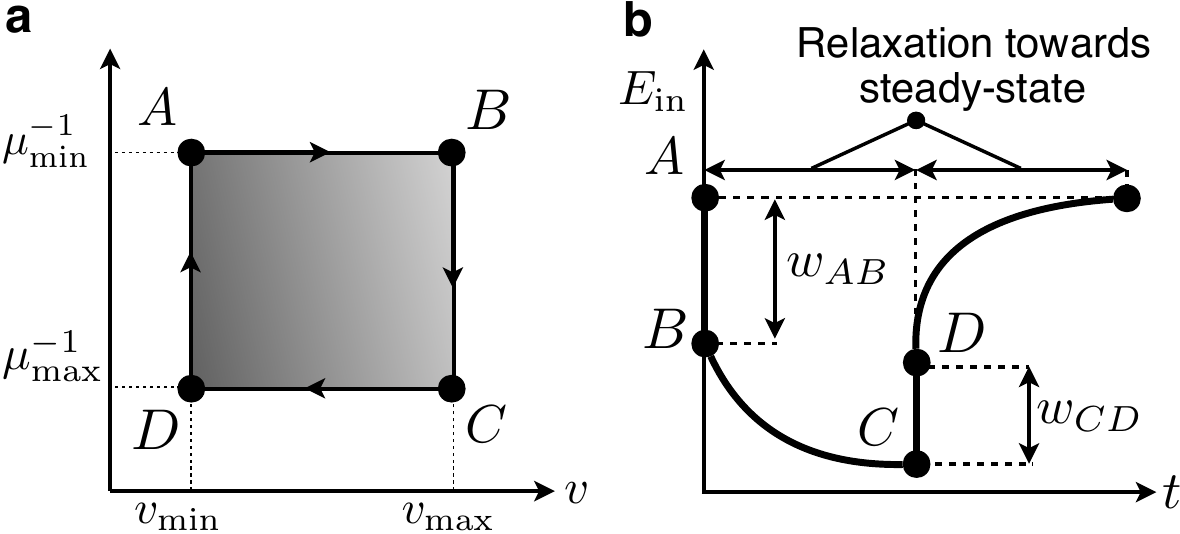}
\caption{{\bf Periodic driving and representative cycle of the internal energy.} {\bf a} Periodic  driving protocol in the Rydberg system. The parameters in the various transformation are varied through sudden quenches. {\bf b} Representative instance of the time-dependent internal energy for the cycle shown in {\bf a}. Sudden jumps in $E_{\rm in}$ are associated with work exchanges through the interparticle interactions (negative in the expansion). Curves with finite derivative represent relaxation after quenches in the laser detuning parameter $\mu$. }
\label{Fig2}
\end{figure} 

The contribution $H_{\rm at}$ does not appear in Eq.~\eqref{H} since the latter represents the Hamiltonian of the system in the interaction picture. The thermodynamic balance however must be formulated in the Schr\"odinger picture \cite{PhysRevA.74.063823}. The unitary connecting Schr\"odinger and interaction picture has a generator proportional to $(\mu+\omega_{\rm at})\sum_k n^{(k)}$, which commutes with $H_{\rm in}^v$. Therefore, after identifying the  internal energy  we can derive the thermodynamic balance using Eq.~\eqref{int-picture}. The choice of the internal energy in Eq.~\eqref{int-energy} derives from a microscopic picture where the atomic chain, i.e., the system proper, is perturbed by the laser and the interaction with a thermal environment. In this setting, the Hamiltonian terms proportional to $\Omega$ and $\mu$ describe the interaction of the system with a laser, while the dissipative term, proportional to $\gamma$,  describes its interaction with an environment. The actual internal energy of the atom system is thus solely associated with the Hamiltonian term in Eq.~\eqref{int-operator}. 

The first law thus reads 
\begin{equation}
\dot{E}_{\rm in}=-f_t^{v_t}\, \dot{v}_t+I_t-J_t\, .
\label{first-law}
\end{equation}
Here,
\begin{equation}
f_t^{v_t}=-\Tr\left[\rho_t\, \partial_v H_{\rm in}^v\right]\big|_{v=v_{t}}\, ,
\label{ft-work}
\end{equation}
is the generalised force associated with the specific volume $v$, which is the only term in the internal energy responsible for  mechanical work in our setup. The input power provided by the laser is  given by 
$$
I_t=i\Tr\left[\rho_t\, [H_{\rm ex}^\mu , H_{\rm in}^v]\right]\big|_{v=v_t}^{\mu=\mu_t}\, ,
$$
where the term $H_{\rm ex}^\mu=H-H_{\rm LG}^v$ corresponds to the laser-atom interaction. Finally,  we have to account for the dissipated heat 
\begin{equation*}
J_t=-\Tr\left[\rho_t\, \mathcal{L}[H_{\rm in}^{v}]\right]\big|_{v=v_t}\, .
\label{heat-flux}
\end{equation*}
Note that, due to the form of $H_{\rm LG}^v$, this quantity is non-negative  ($J_t\ge0$) and vanishes only in the zero-excitation state. The environment is effectively at zero temperature as it is highly unlikely to observe spontaneous excitation of atoms at room temperature. Therefore, the above properties of $J_t$ are sufficient to guarantee consistency with the second law: no heat is extracted from the zero-temperature environment. 

The engine cycle we propose consists of a four-stroke periodic driving involving sudden quenches of the parameters $\mu$ and $v$, as depicted in Fig.~\ref{Fig2}(a). The presence of radiative decay is thereby of fundamental importance. This feature prevents the system from heating up to infinite temperature  \cite{PhysRevX.4.041048,PhysRevE.90.012110,PONTE2015196}, as it would happen for an isolated engine. Instead, the system state approaches a non-trivial asymptotic cycle (see discussion in the Supplementary Information). The actual cycle starts from $A$ with a transformation given by a sudden expansion of the volume $v_{\rm min}\to v_{\rm max}=v_{\rm min}+\Delta v$. During this expansion, work is extracted from the system. Afterwards, the system immediately undergoes a sudden quench $\mu_{\rm min}\to\mu_{\rm max}=\mu_{\rm min}+\Delta \mu$. In this transformation, no work is exchanged, but the system now evolves for a relaxation period, shown in Fig.~\ref{Fig2}(b), reaching a non-equilibrium state with a lower mean interparticle energy. The following sudden compression $v_{\rm max}\to v_{\rm min}$ thus requires less work than is extracted during the expansion. Immediately after the compression, the transformation $\mu_{\rm max}\to\mu_{\rm min}$ takes the system back to its initial state through a second relaxation period. In Fig.~\ref{Fig2}(b) we show a representative cycle of the internal energy following the periodic driving which further highlights the two relaxation periods in the cycle. In the regime depicted in Fig.~\ref{Fig2}(b), the engine delivers positive net output. 

To quantify the net  output of our Rydberg engine, we integrate the first law over a full period obtaining 
$$
w_{\rm net}:=\int dt\, f_t^{v_t}\dot{v}_t=I-J\, ,
$$
where $I=\int dt\, I_t$ is the total input in one cycle and $J=\int J_t$ is the total dissipated heat. For $w_{\rm net}\ge0$, we can thus define the efficiency of the engine as
$$
\eta=\frac{w_{\rm net}}{I}=\frac{w_{\rm net}}{w_{\rm net}+J}\le1\, ,
$$
where the last equality follows from $J_t\ge0$. Adapting these quantities to our driving protocol, we have 
$$
w_{\rm net}=-w_{AB}-w_{CD}\, 
$$
with (see Supplementary Information)
\begin{equation}
\begin{split}
&w_{AB}=\left(\frac{1}{v_{\rm max}}-\frac{1}{v_{\rm min}}\right)\frac{1}{N_{\rm at}}\sum_{k=1}^{N_{\rm at}-1} \langle n^{(k)}n^{(k+1)}\rangle_A\le0\,,\\
&w_{CD}=\left(\frac{1}{v_{\rm min}}-\frac{1}{v_{\rm max}}\right)\frac{1}{N_{\rm at}}\sum_{k=1}^{N_{\rm at}-1} \langle n^{(k)}n^{(k+1)}\rangle_C\ge0\,.
\end{split}
\label{exchanges}
\end{equation}
In the above equations, $\langle \cdot \rangle_{A/C}$ is the expectation value taken over the non-equilibrium steady-state corresponding to the parameters in $A/C$, respectively. To produce net work, one needs to have $|w_{AB}|>|w_{CD}|$ [{\it cf.} Fig.~\ref{Fig2}(b)] which directly translates in $\sum_k\langle n^{(k)}n^{(k+1)}\rangle_A>\sum_k\langle n^{(k)}n^{(k+1)}\rangle_C$. Hence, interactions must be stronger during the expansion. Furthermore, as we are considering infinite relaxation periods, we have $J\to\infty$, since constant dissipation, also known as {\it housekeeping heat}, is required to maintain the non-equilibrium steady-state even if no thermodynamic transformation is performed. For finite relaxation times, where $J_t<\infty$,  it is however possible to have a finite efficiency also in our non-equilibrium setting. \\

\begin{figure}[t]
\centering
\includegraphics[scale=0.5]{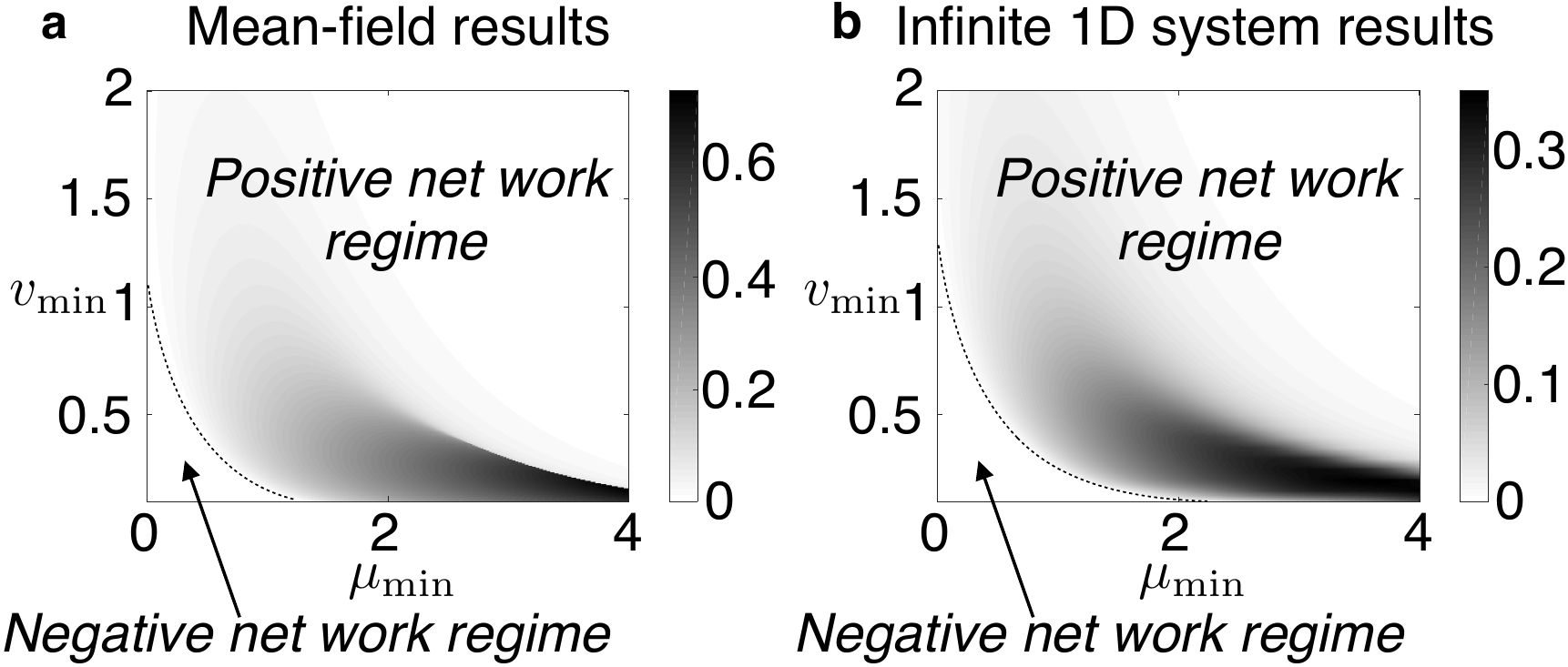}
\caption{{\bf Dissipative infinite 1D Rydberg chain. } Phase diagram for a Rydberg engine. The parameters of the cycle for both plots are $\Omega=1$, $\gamma=1$, $\Delta v=\Delta \mu =1$ and the relaxation time involved in the cycle is infinite. We focus on the stationary cycle. {\bf a} Mean-field. The panel displays a density plot for the work delivered by the engine as a function of $v_{\rm min}$ and $\mu_{\rm min}$. It is indeed possible to have cycles that produce positive net work. {\bf b} Infinite 1D Rydberg atomic system with nearest neighbour interactions. We observe the same qualitative features displayed by mean-field results. From a quantitative point of view, the work that can be extracted in an infinite 1D system differs from the one estimated in the mean-field approximation. In particular, the maximum amount of net work that can be extracted in a 1D system is smaller than  predicted by the mean-field theory.}
\label{Fig4}
\end{figure} 

\noindent {\bf \em Simulation results-- } We now explore numerically the cycle as a function of the chosen values of $v_{\rm min}$ and $\mu_{\rm min}$. We  fix $\Omega=\gamma=1$ and assume that, after quenching $\mu$, the system fully relaxes to its genuine quantum non-equilibrium steady-state.

In Fig.~\ref{Fig4}(a), we show the net work produced by an infinite system within a mean-field approximation. In Fig.~\ref{Fig4}(b), instead, we display numerically exact data for a 1D chain obtained via infinite matrix product algorithms \cite{Vidal2007,paeckel2019timeevolution} (see Methods for details on both techniques). 
First, we observe that it is indeed possible to extract net work from our cycle. Many-body simulations, accounting for  quantum correlations in a nonperturbative way, confirm that this effect is not only a feature of the non-linear mean-field dynamics. In particular, a boundary arises in the phase diagrams of Fig.~\ref{Fig4}(a-b) separating a region where mechanical work can be extracted from the atomic array and a region where the system absorbs energy instead of delivering work. Additional finite-size and finite-relaxation results (see Supplementary Information) show unambiguously that positive output work can be extracted also in finite systems and very importantly also with finite relaxation times, i.e. with a finite efficiency. \\

\noindent {\bf \em  Discussion-- } We have developed a many-body quantum engine that operates intrinsically under non-equilibrium conditions. While this design strategy, in contrast to conventional heat engine cycles, requires additional energy input to keep the working system in a non-equilibrium state ($J_t$ does not vanish even at stationarity), it also allows for a more direct and efficient control of physical parameters, e.g. through external laser fields. Focussing on Rydberg atom ensembles, we have shown that it is possible to generate useful work.   In principle, this output could be extracted through a movable device, e.g.~a mirror interfacing with the Rydberg atomic array, and can thus be considered as actual mechanical work. To verify our theoretical predictions, experiments measuring the spatially resolved density of excitations suffice. These data would allow to estimate density-density correlation functions and thus to infer the produced work. 

In our current considerations, we did not consider the effect such measurement of the position of atoms would have on the thermodynamic cycle and on fluctuations of the generated work. Instead, we have treated the interparticle distance as a classical parameter, which, in the spirit of a Born-Oppenheimer approximation, decouples from the electronic degrees of freedom of the atoms. It would be interesting to derive from first principles a description of our engine which accounts for cross-correlations between position and momentum of the atoms and their electronic state. 

It further remains an open question  whether non-thermalising closed quantum systems, as in the case of many-body localisation \cite{PhysRevLett.116.250401,ABANIN20161,PhysRevLett.114.140401,PhysRevLett.115.030402} or prethermal metastable regimes due to almost conserved charges \cite{kuwahara2016,Abanin2017}, might be exploited to devise an isolated non-equilibrium many-body engine with a non-trivial cycle.

\begin{acknowledgments}
The research leading to these results has received funding  from EPSRC Grants No. EP/M014266/1, EP/N03404X/1, and No. EP/R04340X/1 via the QuantERA project “ERyQSenS”. K.B. is an International Research Fellow of the Japan Society for the Promotion of Science (Fellowship ID: P19026). I.L. acknowledges support from the DFG through SPP 1929 (GiRyd) as well as from the ``Wissenschaftler-R\"uckkehrprogramm GSO/CZS" of the Carl-Zeiss-Stiftung and the German Scholars Organization e.V. We are grateful for access to the University of Nottingham's Augusta HPC service. We also acknowledge the use of Athena at HPC Midlands Plus.
\end{acknowledgments}

\bibliography{BiblioNotesWork}{}

\begin{thebibliography}{60}%
\makeatletter
\providecommand \@ifxundefined [1]{%
 \@ifx{#1\undefined}
}%
\providecommand \@ifnum [1]{%
 \ifnum #1\expandafter \@firstoftwo
 \else \expandafter \@secondoftwo
 \fi
}%
\providecommand \@ifx [1]{%
 \ifx #1\expandafter \@firstoftwo
 \else \expandafter \@secondoftwo
 \fi
}%
\providecommand \natexlab [1]{#1}%
\providecommand \enquote  [1]{``#1''}%
\providecommand \bibnamefont  [1]{#1}%
\providecommand \bibfnamefont [1]{#1}%
\providecommand \citenamefont [1]{#1}%
\providecommand \href@noop [0]{\@secondoftwo}%
\providecommand \href [0]{\begingroup \@sanitize@url \@href}%
\providecommand \@href[1]{\@@startlink{#1}\@@href}%
\providecommand \@@href[1]{\endgroup#1\@@endlink}%
\providecommand \@sanitize@url [0]{\catcode `\\12\catcode `\$12\catcode
  `\&12\catcode `\#12\catcode `\^12\catcode `\_12\catcode `\%12\relax}%
\providecommand \@@startlink[1]{}%
\providecommand \@@endlink[0]{}%
\providecommand \url  [0]{\begingroup\@sanitize@url \@url }%
\providecommand \@url [1]{\endgroup\@href {#1}{\urlprefix }}%
\providecommand \urlprefix  [0]{URL }%
\providecommand \Eprint [0]{\href }%
\providecommand \doibase [0]{http://dx.doi.org/}%
\providecommand \selectlanguage [0]{\@gobble}%
\providecommand \bibinfo  [0]{\@secondoftwo}%
\providecommand \bibfield  [0]{\@secondoftwo}%
\providecommand \translation [1]{[#1]}%
\providecommand \BibitemOpen [0]{}%
\providecommand \bibitemStop [0]{}%
\providecommand \bibitemNoStop [0]{.\EOS\space}%
\providecommand \EOS [0]{\spacefactor3000\relax}%
\providecommand \BibitemShut  [1]{\csname bibitem#1\endcsname}%
\let\auto@bib@innerbib\@empty
\bibitem [{\citenamefont {et~al.}(2015)}]{Martinez:2015aa}%
  \BibitemOpen
  \bibfield  {author} {\bibinfo {author} {\bibfnamefont {I.~A.~Mart{\'\i}nez}\
  \bibnamefont {et~al.}},\ }\bibfield  {title} {\enquote {\bibinfo {title}
  {Brownian carnot engine},}\ }\href {https://doi.org/10.1038/nphys3518}
  {\bibfield  {journal} {\bibinfo  {journal} {Nat. Phys.}\ }\textbf {\bibinfo
  {volume} {12}},\ \bibinfo {pages} {67} (\bibinfo {year} {2015})}\BibitemShut
  {NoStop}%
\bibitem [{\citenamefont {Krishnamurthy}\ \emph {et~al.}(2016)\citenamefont
  {Krishnamurthy}, \citenamefont {Ghosh}, \citenamefont {Chatterji},
  \citenamefont {Ganapathy},\ and\ \citenamefont
  {Sood}}]{Krishnamurthy:2016aa}%
  \BibitemOpen
  \bibfield  {author} {\bibinfo {author} {\bibfnamefont {S.}~\bibnamefont
  {Krishnamurthy}}, \bibinfo {author} {\bibfnamefont {S.}~\bibnamefont
  {Ghosh}}, \bibinfo {author} {\bibfnamefont {D.}~\bibnamefont {Chatterji}},
  \bibinfo {author} {\bibfnamefont {R.}~\bibnamefont {Ganapathy}}, \ and\
  \bibinfo {author} {\bibfnamefont {A.~K.}\ \bibnamefont {Sood}},\ }\bibfield
  {title} {\enquote {\bibinfo {title} {A micrometre-sized heat engine operating
  between bacterial reservoirs},}\ }\href {https://doi.org/10.1038/nphys3870}
  {\bibfield  {journal} {\bibinfo  {journal} {Nat. Phys.}\ }\textbf {\bibinfo
  {volume} {12}},\ \bibinfo {pages} {1134} (\bibinfo {year}
  {2016})}\BibitemShut {NoStop}%
\bibitem [{\citenamefont {et~al.}(2016{\natexlab{a}})}]{Rossnagel325}%
  \BibitemOpen
  \bibfield  {author} {\bibinfo {author} {\bibfnamefont {J.~Ro{\ss}nagel}\
  \bibnamefont {et~al.}},\ }\bibfield  {title} {\enquote {\bibinfo {title} {A
  single-atom heat engine},}\ }\href {\doibase 10.1126/science.aad6320}
  {\bibfield  {journal} {\bibinfo  {journal} {Science}\ }\textbf {\bibinfo
  {volume} {352}},\ \bibinfo {pages} {325--329} (\bibinfo {year}
  {2016}{\natexlab{a}})}\BibitemShut {NoStop}%
\bibitem [{\citenamefont {et~al.}(2018{\natexlab{a}})}]{Josefsson:2018aa}%
  \BibitemOpen
  \bibfield  {author} {\bibinfo {author} {\bibfnamefont {M.~Josefsson}\
  \bibnamefont {et~al.}},\ }\bibfield  {title} {\enquote {\bibinfo {title} {A
  quantum-dot heat engine operating close to the thermodynamic efficiency
  limits},}\ }\href {\doibase 10.1038/s41565-018-0200-5} {\bibfield  {journal}
  {\bibinfo  {journal} {Nat. Nanotechnol.}\ }\textbf {\bibinfo {volume} {13}},\
  \bibinfo {pages} {920--924} (\bibinfo {year}
  {2018}{\natexlab{a}})}\BibitemShut {NoStop}%
\bibitem [{\citenamefont {Schliwa}\ and\ \citenamefont
  {Woehlke}(2003)}]{Schliwa:2003aa}%
  \BibitemOpen
  \bibfield  {author} {\bibinfo {author} {\bibfnamefont {M.}~\bibnamefont
  {Schliwa}}\ and\ \bibinfo {author} {\bibfnamefont {G.}~\bibnamefont
  {Woehlke}},\ }\bibfield  {title} {\enquote {\bibinfo {title} {Molecular
  motors},}\ }\href {\doibase 10.1038/nature01601} {\bibfield  {journal}
  {\bibinfo  {journal} {Nature}\ }\textbf {\bibinfo {volume} {422}},\ \bibinfo
  {pages} {759--765} (\bibinfo {year} {2003})}\BibitemShut {NoStop}%
\bibitem [{\citenamefont {Seifert}(2012)}]{Seifert_2012}%
  \BibitemOpen
  \bibfield  {author} {\bibinfo {author} {\bibfnamefont {U.}~\bibnamefont
  {Seifert}},\ }\bibfield  {title} {\enquote {\bibinfo {title} {Stochastic
  thermodynamics, fluctuation theorems and molecular machines},}\ }\href
  {\doibase 10.1088/0034-4885/75/12/126001} {\bibfield  {journal} {\bibinfo
  {journal} {Rep. Prog. Phys.}\ }\textbf {\bibinfo {volume} {75}},\ \bibinfo
  {pages} {126001} (\bibinfo {year} {2012})}\BibitemShut {NoStop}%
\bibitem [{\citenamefont {Elouard}\ and\ \citenamefont
  {Jordan}(2018)}]{PhysRevLett.120.260601}%
  \BibitemOpen
  \bibfield  {author} {\bibinfo {author} {\bibfnamefont {C.}~\bibnamefont
  {Elouard}}\ and\ \bibinfo {author} {\bibfnamefont {A.~N.}\ \bibnamefont
  {Jordan}},\ }\bibfield  {title} {\enquote {\bibinfo {title} {Efficient
  quantum measurement engines},}\ }\href {\doibase
  10.1103/PhysRevLett.120.260601} {\bibfield  {journal} {\bibinfo  {journal}
  {Phys. Rev. Lett.}\ }\textbf {\bibinfo {volume} {120}},\ \bibinfo {pages}
  {260601} (\bibinfo {year} {2018})}\BibitemShut {NoStop}%
\bibitem [{\citenamefont
  {et~al.}(2019{\natexlab{a}})}]{PhysRevLett.122.110601}%
  \BibitemOpen
  \bibfield  {author} {\bibinfo {author} {\bibfnamefont {J.~Klatzow}\
  \bibnamefont {et~al.}},\ }\bibfield  {title} {\enquote {\bibinfo {title}
  {Experimental demonstration of quantum effects in the operation of
  microscopic heat engines},}\ }\href {\doibase 10.1103/PhysRevLett.122.110601}
  {\bibfield  {journal} {\bibinfo  {journal} {Phys. Rev. Lett.}\ }\textbf
  {\bibinfo {volume} {122}},\ \bibinfo {pages} {110601} (\bibinfo {year}
  {2019}{\natexlab{a}})}\BibitemShut {NoStop}%
\bibitem [{\citenamefont {Uzdin}\ \emph {et~al.}(2015)\citenamefont {Uzdin},
  \citenamefont {Levy},\ and\ \citenamefont {Kosloff}}]{PhysRevX.5.031044}%
  \BibitemOpen
  \bibfield  {author} {\bibinfo {author} {\bibfnamefont {R.}~\bibnamefont
  {Uzdin}}, \bibinfo {author} {\bibfnamefont {A.}~\bibnamefont {Levy}}, \ and\
  \bibinfo {author} {\bibfnamefont {R.}~\bibnamefont {Kosloff}},\ }\bibfield
  {title} {\enquote {\bibinfo {title} {Equivalence of quantum heat machines,
  and quantum-thermodynamic signatures},}\ }\href {\doibase
  10.1103/PhysRevX.5.031044} {\bibfield  {journal} {\bibinfo  {journal} {Phys.
  Rev. X}\ }\textbf {\bibinfo {volume} {5}},\ \bibinfo {pages} {031044}
  (\bibinfo {year} {2015})}\BibitemShut {NoStop}%
\bibitem [{\citenamefont {Ro\ss{}nagel}\ \emph {et~al.}(2014)\citenamefont
  {Ro\ss{}nagel}, \citenamefont {Abah}, \citenamefont {Schmidt-Kaler},
  \citenamefont {Singer},\ and\ \citenamefont {Lutz}}]{PhysRevLett.112.030602}%
  \BibitemOpen
  \bibfield  {author} {\bibinfo {author} {\bibfnamefont {J.}~\bibnamefont
  {Ro\ss{}nagel}}, \bibinfo {author} {\bibfnamefont {O.}~\bibnamefont {Abah}},
  \bibinfo {author} {\bibfnamefont {F.}~\bibnamefont {Schmidt-Kaler}}, \bibinfo
  {author} {\bibfnamefont {K.}~\bibnamefont {Singer}}, \ and\ \bibinfo {author}
  {\bibfnamefont {E.}~\bibnamefont {Lutz}},\ }\bibfield  {title} {\enquote
  {\bibinfo {title} {Nanoscale heat engine beyond the carnot limit},}\ }\href
  {\doibase 10.1103/PhysRevLett.112.030602} {\bibfield  {journal} {\bibinfo
  {journal} {Phys. Rev. Lett.}\ }\textbf {\bibinfo {volume} {112}},\ \bibinfo
  {pages} {030602} (\bibinfo {year} {2014})}\BibitemShut {NoStop}%
\bibitem [{\citenamefont {Jaramillo}\ \emph {et~al.}(2016)\citenamefont
  {Jaramillo}, \citenamefont {Beau},\ and\ \citenamefont {del
  Campo}}]{Jaramillo_2016}%
  \BibitemOpen
  \bibfield  {author} {\bibinfo {author} {\bibfnamefont {J.}~\bibnamefont
  {Jaramillo}}, \bibinfo {author} {\bibfnamefont {M.}~\bibnamefont {Beau}}, \
  and\ \bibinfo {author} {\bibfnamefont {A.}~\bibnamefont {del Campo}},\
  }\bibfield  {title} {\enquote {\bibinfo {title} {Quantum supremacy of
  many-particle thermal machines},}\ }\href {\doibase
  10.1088/1367-2630/18/7/075019} {\bibfield  {journal} {\bibinfo  {journal}
  {New J. Phys.}\ }\textbf {\bibinfo {volume} {18}},\ \bibinfo {pages} {075019}
  (\bibinfo {year} {2016})}\BibitemShut {NoStop}%
\bibitem [{\citenamefont {Halpern}\ \emph {et~al.}(2019)\citenamefont
  {Halpern}, \citenamefont {White}, \citenamefont {Gopalakrishnan},\ and\
  \citenamefont {Refael}}]{PhysRevB.99.024203}%
  \BibitemOpen
  \bibfield  {author} {\bibinfo {author} {\bibfnamefont {N.~Yunger}\
  \bibnamefont {Halpern}}, \bibinfo {author} {\bibfnamefont {C.~D.}\
  \bibnamefont {White}}, \bibinfo {author} {\bibfnamefont {S.}~\bibnamefont
  {Gopalakrishnan}}, \ and\ \bibinfo {author} {\bibfnamefont {G.}~\bibnamefont
  {Refael}},\ }\bibfield  {title} {\enquote {\bibinfo {title} {Quantum engine
  based on many-body localization},}\ }\href {\doibase
  10.1103/PhysRevB.99.024203} {\bibfield  {journal} {\bibinfo  {journal} {Phys.
  Rev. B}\ }\textbf {\bibinfo {volume} {99}},\ \bibinfo {pages} {024203}
  (\bibinfo {year} {2019})}\BibitemShut {NoStop}%
\bibitem [{\citenamefont {Vinjanampathy}\ and\ \citenamefont
  {Anders}(2016)}]{anders}%
  \BibitemOpen
  \bibfield  {author} {\bibinfo {author} {\bibfnamefont {S.}~\bibnamefont
  {Vinjanampathy}}\ and\ \bibinfo {author} {\bibfnamefont {J.}~\bibnamefont
  {Anders}},\ }\bibfield  {title} {\enquote {\bibinfo {title} {Quantum
  thermodynamics},}\ }\href {\doibase 10.1080/00107514.2016.1201896} {\bibfield
   {journal} {\bibinfo  {journal} {Contemp. Phys.}\ }\textbf {\bibinfo {volume}
  {57}},\ \bibinfo {pages} {545--579} (\bibinfo {year} {2016})}\BibitemShut
  {NoStop}%
\bibitem [{\citenamefont {Alicki}\ and\ \citenamefont
  {Kosloff}(2018)}]{Alicki2018}%
  \BibitemOpen
  \bibfield  {author} {\bibinfo {author} {\bibfnamefont {R.}~\bibnamefont
  {Alicki}}\ and\ \bibinfo {author} {\bibfnamefont {R.}~\bibnamefont
  {Kosloff}},\ }\enquote {\bibinfo {title} {Introduction to quantum
  thermodynamics: History and prospects},}\ in\ \href {\doibase
  10.1007/978-3-319-99046-0_1} {\emph {\bibinfo {booktitle} {Thermodynamics in
  the Quantum Regime: Fundamental Aspects and New Directions}}},\ \bibinfo
  {editor} {edited by\ \bibinfo {editor} {\bibfnamefont {Felix}\ \bibnamefont
  {Binder}}, \bibinfo {editor} {\bibfnamefont {Luis~A.}\ \bibnamefont
  {Correa}}, \bibinfo {editor} {\bibfnamefont {Christian}\ \bibnamefont
  {Gogolin}}, \bibinfo {editor} {\bibfnamefont {Janet}\ \bibnamefont {Anders}},
  \ and\ \bibinfo {editor} {\bibfnamefont {Gerardo}\ \bibnamefont {Adesso}}}\
  (\bibinfo  {publisher} {Springer International Publishing},\ \bibinfo
  {address} {Cham},\ \bibinfo {year} {2018})\ pp.\ \bibinfo {pages}
  {1--33}\BibitemShut {NoStop}%
\bibitem [{\citenamefont {Deffner}\ and\ \citenamefont
  {Campbell}(2019)}]{Deffner2019}%
  \BibitemOpen
  \bibfield  {author} {\bibinfo {author} {\bibfnamefont {S.}~\bibnamefont
  {Deffner}}\ and\ \bibinfo {author} {\bibfnamefont {S.}~\bibnamefont
  {Campbell}},\ }\href {\doibase 10.1088/2053-2571/ab21c6} {\emph {\bibinfo
  {title} {Quantum Thermodynamics}}},\ 2053-2571\ (\bibinfo  {publisher}
  {Morgan \& Claypool Publishers},\ \bibinfo {year} {2019})\BibitemShut
  {NoStop}%
\bibitem [{\citenamefont {Fialko}\ and\ \citenamefont
  {Hallwood}(2012)}]{PhysRevLett.108.085303}%
  \BibitemOpen
  \bibfield  {author} {\bibinfo {author} {\bibfnamefont {O.}~\bibnamefont
  {Fialko}}\ and\ \bibinfo {author} {\bibfnamefont {D.~W.}\ \bibnamefont
  {Hallwood}},\ }\bibfield  {title} {\enquote {\bibinfo {title} {Isolated
  quantum heat engine},}\ }\href {\doibase 10.1103/PhysRevLett.108.085303}
  {\bibfield  {journal} {\bibinfo  {journal} {Phys. Rev. Lett.}\ }\textbf
  {\bibinfo {volume} {108}},\ \bibinfo {pages} {085303} (\bibinfo {year}
  {2012})}\BibitemShut {NoStop}%
\bibitem [{\citenamefont {Chen}\ \emph {et~al.}(2019)\citenamefont {Chen},
  \citenamefont {Watanabe}, \citenamefont {Yu}, \citenamefont {Guan},\ and\
  \citenamefont {del Campo}}]{Chen:2019aa}%
  \BibitemOpen
  \bibfield  {author} {\bibinfo {author} {\bibfnamefont {Y.-Y.}\ \bibnamefont
  {Chen}}, \bibinfo {author} {\bibfnamefont {G.}~\bibnamefont {Watanabe}},
  \bibinfo {author} {\bibfnamefont {Y.-C.}\ \bibnamefont {Yu}}, \bibinfo
  {author} {\bibfnamefont {X.-W.}\ \bibnamefont {Guan}}, \ and\ \bibinfo
  {author} {\bibfnamefont {A.}~\bibnamefont {del Campo}},\ }\bibfield  {title}
  {\enquote {\bibinfo {title} {An interaction-driven many-particle quantum heat
  engine and its universal behavior},}\ }\href {\doibase
  10.1038/s41534-019-0204-5} {\bibfield  {journal} {\bibinfo  {journal} {Npj
  Quantum Inf.}\ }\textbf {\bibinfo {volume} {5}},\ \bibinfo {pages} {88}
  (\bibinfo {year} {2019})}\BibitemShut {NoStop}%
\bibitem [{\citenamefont {Bloch}(2005)}]{Bloch:2005aa}%
  \BibitemOpen
  \bibfield  {author} {\bibinfo {author} {\bibfnamefont {I.}~\bibnamefont
  {Bloch}},\ }\bibfield  {title} {\enquote {\bibinfo {title} {Ultracold quantum
  gases in optical lattices},}\ }\href {\doibase 10.1038/nphys138} {\bibfield
  {journal} {\bibinfo  {journal} {Nat. Phys.}\ }\textbf {\bibinfo {volume}
  {1}},\ \bibinfo {pages} {23--30} (\bibinfo {year} {2005})}\BibitemShut
  {NoStop}%
\bibitem [{\citenamefont {Anderson}\ \emph {et~al.}(2011)\citenamefont
  {Anderson}, \citenamefont {Younge},\ and\ \citenamefont
  {Raithel}}]{PhysRevLett.107.263001}%
  \BibitemOpen
  \bibfield  {author} {\bibinfo {author} {\bibfnamefont {S.~E.}\ \bibnamefont
  {Anderson}}, \bibinfo {author} {\bibfnamefont {K.~C.}\ \bibnamefont
  {Younge}}, \ and\ \bibinfo {author} {\bibfnamefont {G.}~\bibnamefont
  {Raithel}},\ }\bibfield  {title} {\enquote {\bibinfo {title} {Trapping
  rydberg atoms in an optical lattice},}\ }\href {\doibase
  10.1103/PhysRevLett.107.263001} {\bibfield  {journal} {\bibinfo  {journal}
  {Phys. Rev. Lett.}\ }\textbf {\bibinfo {volume} {107}},\ \bibinfo {pages}
  {263001} (\bibinfo {year} {2011})}\BibitemShut {NoStop}%
\bibitem [{\citenamefont {Bloch}\ \emph {et~al.}(2008)\citenamefont {Bloch},
  \citenamefont {Dalibard},\ and\ \citenamefont {Zwerger}}]{RevModPhys.80.885}%
  \BibitemOpen
  \bibfield  {author} {\bibinfo {author} {\bibfnamefont {I.}~\bibnamefont
  {Bloch}}, \bibinfo {author} {\bibfnamefont {J.}~\bibnamefont {Dalibard}}, \
  and\ \bibinfo {author} {\bibfnamefont {W.}~\bibnamefont {Zwerger}},\
  }\bibfield  {title} {\enquote {\bibinfo {title} {Many-body physics with
  ultracold gases},}\ }\href {\doibase 10.1103/RevModPhys.80.885} {\bibfield
  {journal} {\bibinfo  {journal} {Rev. Mod. Phys.}\ }\textbf {\bibinfo {volume}
  {80}},\ \bibinfo {pages} {885--964} (\bibinfo {year} {2008})}\BibitemShut
  {NoStop}%
\bibitem [{\citenamefont {Saffman}\ \emph {et~al.}(2010)\citenamefont
  {Saffman}, \citenamefont {Walker},\ and\ \citenamefont
  {M\o{}lmer}}]{RevModPhys.82.2313}%
  \BibitemOpen
  \bibfield  {author} {\bibinfo {author} {\bibfnamefont {M.}~\bibnamefont
  {Saffman}}, \bibinfo {author} {\bibfnamefont {T.~G.}\ \bibnamefont {Walker}},
  \ and\ \bibinfo {author} {\bibfnamefont {K.}~\bibnamefont {M\o{}lmer}},\
  }\bibfield  {title} {\enquote {\bibinfo {title} {Quantum information with
  rydberg atoms},}\ }\href {\doibase 10.1103/RevModPhys.82.2313} {\bibfield
  {journal} {\bibinfo  {journal} {Rev. Mod. Phys.}\ }\textbf {\bibinfo {volume}
  {82}},\ \bibinfo {pages} {2313--2363} (\bibinfo {year} {2010})}\BibitemShut
  {NoStop}%
\bibitem [{\citenamefont {et~al.}(2016{\natexlab{b}})}]{Endres1024}%
  \BibitemOpen
  \bibfield  {author} {\bibinfo {author} {\bibfnamefont {M.~Endres}\
  \bibnamefont {et~al.}},\ }\bibfield  {title} {\enquote {\bibinfo {title}
  {Atom-by-atom assembly of defect-free one-dimensional cold atom arrays},}\
  }\href {\doibase 10.1126/science.aah3752} {\bibfield  {journal} {\bibinfo
  {journal} {Science}\ }\textbf {\bibinfo {volume} {354}},\ \bibinfo {pages}
  {1024--1027} (\bibinfo {year} {2016}{\natexlab{b}})}\BibitemShut {NoStop}%
\bibitem [{\citenamefont {Barredo}\ \emph {et~al.}(2016)\citenamefont
  {Barredo}, \citenamefont {de~L{\'e}s{\'e}leuc}, \citenamefont {Lienhard},
  \citenamefont {Lahaye},\ and\ \citenamefont {Browaeys}}]{Barredo1021}%
  \BibitemOpen
  \bibfield  {author} {\bibinfo {author} {\bibfnamefont {D.}~\bibnamefont
  {Barredo}}, \bibinfo {author} {\bibfnamefont {S.}~\bibnamefont
  {de~L{\'e}s{\'e}leuc}}, \bibinfo {author} {\bibfnamefont {V.}~\bibnamefont
  {Lienhard}}, \bibinfo {author} {\bibfnamefont {T.}~\bibnamefont {Lahaye}}, \
  and\ \bibinfo {author} {\bibfnamefont {A.}~\bibnamefont {Browaeys}},\
  }\bibfield  {title} {\enquote {\bibinfo {title} {An atom-by-atom assembler of
  defect-free arbitrary two-dimensional atomic arrays},}\ }\href {\doibase
  10.1126/science.aah3778} {\bibfield  {journal} {\bibinfo  {journal}
  {Science}\ }\textbf {\bibinfo {volume} {354}},\ \bibinfo {pages} {1021--1023}
  (\bibinfo {year} {2016})}\BibitemShut {NoStop}%
\bibitem [{\citenamefont {et~al.}(2016{\natexlab{c}})}]{Labuhn:2016aa}%
  \BibitemOpen
  \bibfield  {author} {\bibinfo {author} {\bibfnamefont {H.~Labuhn}\
  \bibnamefont {et~al.}},\ }\bibfield  {title} {\enquote {\bibinfo {title}
  {Tunable two-dimensional arrays of single rydberg atoms for realizing quantum
  ising models},}\ }\href {https://doi.org/10.1038/nature18274} {\bibfield
  {journal} {\bibinfo  {journal} {Nature}\ }\textbf {\bibinfo {volume} {534}},\
  \bibinfo {pages} {667} (\bibinfo {year} {2016}{\natexlab{c}})}\BibitemShut
  {NoStop}%
\bibitem [{\citenamefont
  {et~al.}(2017{\natexlab{a}})}]{PhysRevLett.118.063606}%
  \BibitemOpen
  \bibfield  {author} {\bibinfo {author} {\bibfnamefont {M.~Marcuzzi}\
  \bibnamefont {et~al.}},\ }\bibfield  {title} {\enquote {\bibinfo {title}
  {Facilitation dynamics and localization phenomena in rydberg lattice gases
  with position disorder},}\ }\href {\doibase 10.1103/PhysRevLett.118.063606}
  {\bibfield  {journal} {\bibinfo  {journal} {Phys. Rev. Lett.}\ }\textbf
  {\bibinfo {volume} {118}},\ \bibinfo {pages} {063606} (\bibinfo {year}
  {2017}{\natexlab{a}})}\BibitemShut {NoStop}%
\bibitem [{\citenamefont {et~al.}(2017{\natexlab{b}})}]{Bernien:2017aa}%
  \BibitemOpen
  \bibfield  {author} {\bibinfo {author} {\bibfnamefont {H.~Bernien}\
  \bibnamefont {et~al.}},\ }\bibfield  {title} {\enquote {\bibinfo {title}
  {Probing many-body dynamics on a 51-atom quantum simulator},}\ }\href
  {https://doi.org/10.1038/nature24622} {\bibfield  {journal} {\bibinfo
  {journal} {Nature}\ }\textbf {\bibinfo {volume} {551}},\ \bibinfo {pages}
  {579} (\bibinfo {year} {2017}{\natexlab{b}})}\BibitemShut {NoStop}%
\bibitem [{\citenamefont {et~al.}(2018{\natexlab{b}})}]{PhysRevX.8.021070}%
  \BibitemOpen
  \bibfield  {author} {\bibinfo {author} {\bibfnamefont {V.~Lienhard}\
  \bibnamefont {et~al.}},\ }\bibfield  {title} {\enquote {\bibinfo {title}
  {Observing the space- and time-dependent growth of correlations in
  dynamically tuned synthetic ising models with antiferromagnetic
  interactions},}\ }\href {\doibase 10.1103/PhysRevX.8.021070} {\bibfield
  {journal} {\bibinfo  {journal} {Phys. Rev. X}\ }\textbf {\bibinfo {volume}
  {8}},\ \bibinfo {pages} {021070} (\bibinfo {year}
  {2018}{\natexlab{b}})}\BibitemShut {NoStop}%
\bibitem [{\citenamefont {et~al.}(2018{\natexlab{c}})}]{PhysRevX.8.041055}%
  \BibitemOpen
  \bibfield  {author} {\bibinfo {author} {\bibfnamefont {A.~Cooper}\
  \bibnamefont {et~al.}},\ }\bibfield  {title} {\enquote {\bibinfo {title}
  {Alkaline-earth atoms in optical tweezers},}\ }\href {\doibase
  10.1103/PhysRevX.8.041055} {\bibfield  {journal} {\bibinfo  {journal} {Phys.
  Rev. X}\ }\textbf {\bibinfo {volume} {8}},\ \bibinfo {pages} {041055}
  (\bibinfo {year} {2018}{\natexlab{c}})}\BibitemShut {NoStop}%
\bibitem [{\citenamefont {Niedenzu}\ and\ \citenamefont
  {Kurizki}(2018)}]{Niedenzu_2018}%
  \BibitemOpen
  \bibfield  {author} {\bibinfo {author} {\bibfnamefont {W.}~\bibnamefont
  {Niedenzu}}\ and\ \bibinfo {author} {\bibfnamefont {G.}~\bibnamefont
  {Kurizki}},\ }\bibfield  {title} {\enquote {\bibinfo {title} {Cooperative
  many-body enhancement of quantum thermal machine power},}\ }\href {\doibase
  10.1088/1367-2630/aaed55} {\bibfield  {journal} {\bibinfo  {journal} {New. J.
  Phys.}\ }\textbf {\bibinfo {volume} {20}},\ \bibinfo {pages} {113038}
  (\bibinfo {year} {2018})}\BibitemShut {NoStop}%
\bibitem [{\citenamefont {Li}\ \emph {et~al.}(2018)\citenamefont {Li},
  \citenamefont {Fogarty}, \citenamefont {Campbell}, \citenamefont {Chen},\
  and\ \citenamefont {Busch}}]{Li_2018}%
  \BibitemOpen
  \bibfield  {author} {\bibinfo {author} {\bibfnamefont {J.}~\bibnamefont
  {Li}}, \bibinfo {author} {\bibfnamefont {T.}~\bibnamefont {Fogarty}},
  \bibinfo {author} {\bibfnamefont {S.}~\bibnamefont {Campbell}}, \bibinfo
  {author} {\bibfnamefont {X.}~\bibnamefont {Chen}}, \ and\ \bibinfo {author}
  {\bibfnamefont {T.}~\bibnamefont {Busch}},\ }\bibfield  {title} {\enquote
  {\bibinfo {title} {An efficient nonlinear feshbach engine},}\ }\href
  {\doibase 10.1088/1367-2630/aa9cd8} {\bibfield  {journal} {\bibinfo
  {journal} {New J. Phys.}\ }\textbf {\bibinfo {volume} {20}},\ \bibinfo
  {pages} {015005} (\bibinfo {year} {2018})}\BibitemShut {NoStop}%
\bibitem [{\citenamefont {et~al.}(2012)}]{Schaus:2012aa}%
  \BibitemOpen
  \bibfield  {author} {\bibinfo {author} {\bibfnamefont {P.~Schau{\ss}}\
  \bibnamefont {et~al.}},\ }\bibfield  {title} {\enquote {\bibinfo {title}
  {Observation of spatially ordered structures in a two-dimensional rydberg
  gas},}\ }\href {https://doi.org/10.1038/nature11596} {\bibfield  {journal}
  {\bibinfo  {journal} {Nature}\ }\textbf {\bibinfo {volume} {491}},\ \bibinfo
  {pages} {87} (\bibinfo {year} {2012})}\BibitemShut {NoStop}%
\bibitem [{\citenamefont {Lee}\ \emph {et~al.}(2012)\citenamefont {Lee},
  \citenamefont {H\"affner},\ and\ \citenamefont
  {Cross}}]{PhysRevLett.108.023602}%
  \BibitemOpen
  \bibfield  {author} {\bibinfo {author} {\bibfnamefont {T.~E.}\ \bibnamefont
  {Lee}}, \bibinfo {author} {\bibfnamefont {H.}~\bibnamefont {H\"affner}}, \
  and\ \bibinfo {author} {\bibfnamefont {M.~C.}\ \bibnamefont {Cross}},\
  }\bibfield  {title} {\enquote {\bibinfo {title} {Collective quantum jumps of
  rydberg atoms},}\ }\href {\doibase 10.1103/PhysRevLett.108.023602} {\bibfield
   {journal} {\bibinfo  {journal} {Phys. Rev. Lett.}\ }\textbf {\bibinfo
  {volume} {108}},\ \bibinfo {pages} {023602} (\bibinfo {year}
  {2012})}\BibitemShut {NoStop}%
\bibitem [{\citenamefont {Lindblad}(1976)}]{lindblad76a}%
  \BibitemOpen
  \bibfield  {author} {\bibinfo {author} {\bibfnamefont {G.}~\bibnamefont
  {Lindblad}},\ }\bibfield  {title} {\enquote {\bibinfo {title} {Generators of
  quantum dynamical semigroups},}\ }\href@noop {} {\bibfield  {journal}
  {\bibinfo  {journal} {Commun. Math. Phys.}\ }\textbf {\bibinfo {volume}
  {48}},\ \bibinfo {pages} {119--130} (\bibinfo {year} {1976})}\BibitemShut
  {NoStop}%
\bibitem [{\citenamefont {Gorini}\ \emph {et~al.}(1976)\citenamefont {Gorini},
  \citenamefont {Kossakowski},\ and\ \citenamefont {Sudarshan}}]{gorini1976}%
  \BibitemOpen
  \bibfield  {author} {\bibinfo {author} {\bibfnamefont {V.}~\bibnamefont
  {Gorini}}, \bibinfo {author} {\bibfnamefont {A.}~\bibnamefont {Kossakowski}},
  \ and\ \bibinfo {author} {\bibfnamefont {E.~C.~G.}\ \bibnamefont
  {Sudarshan}},\ }\bibfield  {title} {\enquote {\bibinfo {title} {Completely
  positive dynamical semigroups of n-level systems},}\ }\href@noop {}
  {\bibfield  {journal} {\bibinfo  {journal} {J. Math. Phys.}\ }\textbf
  {\bibinfo {volume} {17}},\ \bibinfo {pages} {821--825} (\bibinfo {year}
  {1976})}\BibitemShut {NoStop}%
\bibitem [{\citenamefont {Levy}\ and\ \citenamefont
  {Kosloff}(2014)}]{Levy_2014}%
  \BibitemOpen
  \bibfield  {author} {\bibinfo {author} {\bibfnamefont {A.}~\bibnamefont
  {Levy}}\ and\ \bibinfo {author} {\bibfnamefont {R.}~\bibnamefont {Kosloff}},\
  }\bibfield  {title} {\enquote {\bibinfo {title} {The local approach to
  quantum transport may violate the second law of thermodynamics},}\ }\href
  {\doibase 10.1209/0295-5075/107/20004} {\bibfield  {journal} {\bibinfo
  {journal} {EPL}\ }\textbf {\bibinfo {volume} {107}},\ \bibinfo {pages}
  {20004} (\bibinfo {year} {2014})}\BibitemShut {NoStop}%
\bibitem [{\citenamefont {Barra}(2015)}]{Barra:2015aa}%
  \BibitemOpen
  \bibfield  {author} {\bibinfo {author} {\bibfnamefont {F.}~\bibnamefont
  {Barra}},\ }\bibfield  {title} {\enquote {\bibinfo {title} {The thermodynamic
  cost of driving quantum systems by their boundaries},}\ }\href
  {https://doi.org/10.1038/srep14873} {\bibfield  {journal} {\bibinfo
  {journal} {Sci. Rep.}\ }\textbf {\bibinfo {volume} {5}},\ \bibinfo {pages}
  {14873} (\bibinfo {year} {2015})}\BibitemShut {NoStop}%
\bibitem [{\citenamefont {Brandner}\ and\ \citenamefont
  {Seifert}(2016)}]{PhysRevE.93.062134}%
  \BibitemOpen
  \bibfield  {author} {\bibinfo {author} {\bibfnamefont {K.}~\bibnamefont
  {Brandner}}\ and\ \bibinfo {author} {\bibfnamefont {U.}~\bibnamefont
  {Seifert}},\ }\bibfield  {title} {\enquote {\bibinfo {title} {Periodic
  thermodynamics of open quantum systems},}\ }\href {\doibase
  10.1103/PhysRevE.93.062134} {\bibfield  {journal} {\bibinfo  {journal} {Phys.
  Rev. E}\ }\textbf {\bibinfo {volume} {93}},\ \bibinfo {pages} {062134}
  (\bibinfo {year} {2016})}\BibitemShut {NoStop}%
\bibitem [{\citenamefont {Stockburger}\ and\ \citenamefont
  {Motz}(2017)}]{Stockburger2017}%
  \BibitemOpen
  \bibfield  {author} {\bibinfo {author} {\bibfnamefont {J.~T.}\ \bibnamefont
  {Stockburger}}\ and\ \bibinfo {author} {\bibfnamefont {T.}~\bibnamefont
  {Motz}},\ }\bibfield  {title} {\enquote {\bibinfo {title} {Thermodynamic
  deficiencies of some simple lindblad operators},}\ }\href {\doibase
  10.1002/prop.201600067} {\bibfield  {journal} {\bibinfo  {journal} {Fortschr.
  Phys.}\ }\textbf {\bibinfo {volume} {65}},\ \bibinfo {pages} {1600067}
  (\bibinfo {year} {2017})}\BibitemShut {NoStop}%
\bibitem [{\citenamefont {Boukobza}\ and\ \citenamefont
  {Tannor}(2006)}]{PhysRevA.74.063823}%
  \BibitemOpen
  \bibfield  {author} {\bibinfo {author} {\bibfnamefont {E.}~\bibnamefont
  {Boukobza}}\ and\ \bibinfo {author} {\bibfnamefont {D.~J.}\ \bibnamefont
  {Tannor}},\ }\bibfield  {title} {\enquote {\bibinfo {title} {Thermodynamics
  of bipartite systems: Application to light-matter interactions},}\ }\href
  {\doibase 10.1103/PhysRevA.74.063823} {\bibfield  {journal} {\bibinfo
  {journal} {Phys. Rev. A}\ }\textbf {\bibinfo {volume} {74}},\ \bibinfo
  {pages} {063823} (\bibinfo {year} {2006})}\BibitemShut {NoStop}%
\bibitem [{\citenamefont {D'Alessio}\ and\ \citenamefont
  {Rigol}(2014)}]{PhysRevX.4.041048}%
  \BibitemOpen
  \bibfield  {author} {\bibinfo {author} {\bibfnamefont {L.}~\bibnamefont
  {D'Alessio}}\ and\ \bibinfo {author} {\bibfnamefont {M.}~\bibnamefont
  {Rigol}},\ }\bibfield  {title} {\enquote {\bibinfo {title} {Long-time
  behavior of isolated periodically driven interacting lattice systems},}\
  }\href {\doibase 10.1103/PhysRevX.4.041048} {\bibfield  {journal} {\bibinfo
  {journal} {Phys. Rev. X}\ }\textbf {\bibinfo {volume} {4}},\ \bibinfo {pages}
  {041048} (\bibinfo {year} {2014})}\BibitemShut {NoStop}%
\bibitem [{\citenamefont {Lazarides}\ \emph {et~al.}(2014)\citenamefont
  {Lazarides}, \citenamefont {Das},\ and\ \citenamefont
  {Moessner}}]{PhysRevE.90.012110}%
  \BibitemOpen
  \bibfield  {author} {\bibinfo {author} {\bibfnamefont {A.}~\bibnamefont
  {Lazarides}}, \bibinfo {author} {\bibfnamefont {A.}~\bibnamefont {Das}}, \
  and\ \bibinfo {author} {\bibfnamefont {R.}~\bibnamefont {Moessner}},\
  }\bibfield  {title} {\enquote {\bibinfo {title} {Equilibrium states of
  generic quantum systems subject to periodic driving},}\ }\href {\doibase
  10.1103/PhysRevE.90.012110} {\bibfield  {journal} {\bibinfo  {journal} {Phys.
  Rev. E}\ }\textbf {\bibinfo {volume} {90}},\ \bibinfo {pages} {012110}
  (\bibinfo {year} {2014})}\BibitemShut {NoStop}%
\bibitem [{\citenamefont {Ponte}\ \emph
  {et~al.}(2015{\natexlab{a}})\citenamefont {Ponte}, \citenamefont {Anushya},
  \citenamefont {Papi{\'c}},\ and\ \citenamefont {Abanin}}]{PONTE2015196}%
  \BibitemOpen
  \bibfield  {author} {\bibinfo {author} {\bibfnamefont {P.}~\bibnamefont
  {Ponte}}, \bibinfo {author} {\bibfnamefont {C.}~\bibnamefont {Anushya}},
  \bibinfo {author} {\bibfnamefont {Z.}~\bibnamefont {Papi{\'c}}}, \ and\
  \bibinfo {author} {\bibfnamefont {D.~A.}\ \bibnamefont {Abanin}},\ }\bibfield
   {title} {\enquote {\bibinfo {title} {Periodically driven ergodic and
  many-body localized quantum systems},}\ }\href {\doibase
  https://doi.org/10.1016/j.aop.2014.11.008} {\bibfield  {journal} {\bibinfo
  {journal} {Ann. Phys.}\ }\textbf {\bibinfo {volume} {353}},\ \bibinfo {pages}
  {196 -- 204} (\bibinfo {year} {2015}{\natexlab{a}})}\BibitemShut {NoStop}%
\bibitem [{\citenamefont {Vidal}(2007)}]{Vidal2007}%
  \BibitemOpen
  \bibfield  {author} {\bibinfo {author} {\bibfnamefont {G.}~\bibnamefont
  {Vidal}},\ }\bibfield  {title} {\enquote {\bibinfo {title} {Classical
  simulation of infinite-size quantum lattice systems in one spatial
  dimension},}\ }\href {\doibase 10.1103/PhysRevLett.98.070201} {\bibfield
  {journal} {\bibinfo  {journal} {Phys. Rev. Lett.}\ }\textbf {\bibinfo
  {volume} {98}},\ \bibinfo {pages} {070201} (\bibinfo {year}
  {2007})}\BibitemShut {NoStop}%
\bibitem [{\citenamefont
  {et~al.}(2019{\natexlab{b}})}]{paeckel2019timeevolution}%
  \BibitemOpen
  \bibfield  {author} {\bibinfo {author} {\bibfnamefont {S.~Paeckel}\
  \bibnamefont {et~al.}},\ }\href@noop {} {\enquote {\bibinfo {title}
  {Time-evolution methods for matrix-product states},}\ } (\bibinfo {year}
  {2019}{\natexlab{b}}),\ \Eprint {http://arxiv.org/abs/1901.05824}
  {arXiv:1901.05824 [cond-mat.str-el]} \BibitemShut {NoStop}%
\bibitem [{\citenamefont {Khemani}\ \emph {et~al.}(2016)\citenamefont
  {Khemani}, \citenamefont {Lazarides}, \citenamefont {Moessner},\ and\
  \citenamefont {Sondhi}}]{PhysRevLett.116.250401}%
  \BibitemOpen
  \bibfield  {author} {\bibinfo {author} {\bibfnamefont {V.}~\bibnamefont
  {Khemani}}, \bibinfo {author} {\bibfnamefont {A.}~\bibnamefont {Lazarides}},
  \bibinfo {author} {\bibfnamefont {R.}~\bibnamefont {Moessner}}, \ and\
  \bibinfo {author} {\bibfnamefont {S.~L.}\ \bibnamefont {Sondhi}},\ }\bibfield
   {title} {\enquote {\bibinfo {title} {Phase structure of driven quantum
  systems},}\ }\href {\doibase 10.1103/PhysRevLett.116.250401} {\bibfield
  {journal} {\bibinfo  {journal} {Phys. Rev. Lett.}\ }\textbf {\bibinfo
  {volume} {116}},\ \bibinfo {pages} {250401} (\bibinfo {year}
  {2016})}\BibitemShut {NoStop}%
\bibitem [{\citenamefont {Abanin}\ \emph {et~al.}(2016)\citenamefont {Abanin},
  \citenamefont {Roeck},\ and\ \citenamefont {Huveneers}}]{ABANIN20161}%
  \BibitemOpen
  \bibfield  {author} {\bibinfo {author} {\bibfnamefont {D.~A.}\ \bibnamefont
  {Abanin}}, \bibinfo {author} {\bibfnamefont {W.~De}\ \bibnamefont {Roeck}}, \
  and\ \bibinfo {author} {\bibfnamefont {F.}~\bibnamefont {Huveneers}},\
  }\bibfield  {title} {\enquote {\bibinfo {title} {Theory of many-body
  localization in periodically driven systems},}\ }\href {\doibase
  https://doi.org/10.1016/j.aop.2016.03.010} {\bibfield  {journal} {\bibinfo
  {journal} {Ann. Phys.}\ }\textbf {\bibinfo {volume} {372}},\ \bibinfo {pages}
  {1 -- 11} (\bibinfo {year} {2016})}\BibitemShut {NoStop}%
\bibitem [{\citenamefont {Ponte}\ \emph
  {et~al.}(2015{\natexlab{b}})\citenamefont {Ponte}, \citenamefont {Papic},
  \citenamefont {Huveneers},\ and\ \citenamefont
  {Abanin}}]{PhysRevLett.114.140401}%
  \BibitemOpen
  \bibfield  {author} {\bibinfo {author} {\bibfnamefont {P.}~\bibnamefont
  {Ponte}}, \bibinfo {author} {\bibfnamefont {Z.}~\bibnamefont {Papic}},
  \bibinfo {author} {\bibfnamefont {F.}~\bibnamefont {Huveneers}}, \ and\
  \bibinfo {author} {\bibfnamefont {D.~A.}\ \bibnamefont {Abanin}},\ }\bibfield
   {title} {\enquote {\bibinfo {title} {Many-body localization in periodically
  driven systems},}\ }\href {\doibase 10.1103/PhysRevLett.114.140401}
  {\bibfield  {journal} {\bibinfo  {journal} {Phys. Rev. Lett.}\ }\textbf
  {\bibinfo {volume} {114}},\ \bibinfo {pages} {140401} (\bibinfo {year}
  {2015}{\natexlab{b}})}\BibitemShut {NoStop}%
\bibitem [{\citenamefont {Lazarides}\ \emph {et~al.}(2015)\citenamefont
  {Lazarides}, \citenamefont {Das},\ and\ \citenamefont
  {Moessner}}]{PhysRevLett.115.030402}%
  \BibitemOpen
  \bibfield  {author} {\bibinfo {author} {\bibfnamefont {A.}~\bibnamefont
  {Lazarides}}, \bibinfo {author} {\bibfnamefont {A.}~\bibnamefont {Das}}, \
  and\ \bibinfo {author} {\bibfnamefont {R.}~\bibnamefont {Moessner}},\
  }\bibfield  {title} {\enquote {\bibinfo {title} {Fate of many-body
  localization under periodic driving},}\ }\href {\doibase
  10.1103/PhysRevLett.115.030402} {\bibfield  {journal} {\bibinfo  {journal}
  {Phys. Rev. Lett.}\ }\textbf {\bibinfo {volume} {115}},\ \bibinfo {pages}
  {030402} (\bibinfo {year} {2015})}\BibitemShut {NoStop}%
\bibitem [{\citenamefont {Kuwahara}\ \emph {et~al.}(2016)\citenamefont
  {Kuwahara}, \citenamefont {Mori},\ and\ \citenamefont
  {Saitou}}]{kuwahara2016}%
  \BibitemOpen
  \bibfield  {author} {\bibinfo {author} {\bibfnamefont {T.}~\bibnamefont
  {Kuwahara}}, \bibinfo {author} {\bibfnamefont {T.}~\bibnamefont {Mori}}, \
  and\ \bibinfo {author} {\bibfnamefont {K.}~\bibnamefont {Saitou}},\
  }\bibfield  {title} {\enquote {\bibinfo {title} {Floquet-magnus theory and
  generic transient dynamics in periodically driven many-body quantum
  systems},}\ }\href {\doibase 10.1016/j.aop.2016.01.012} {\bibfield  {journal}
  {\bibinfo  {journal} {Ann. Phys.}\ }\textbf {\bibinfo {volume} {367}},\
  \bibinfo {pages} {96--124} (\bibinfo {year} {2016})}\BibitemShut {NoStop}%
\bibitem [{\citenamefont {Abanin}\ \emph {et~al.}(2017)\citenamefont {Abanin},
  \citenamefont {Roeck}, \citenamefont {Ho},\ and\ \citenamefont
  {Huveneers}}]{Abanin2017}%
  \BibitemOpen
  \bibfield  {author} {\bibinfo {author} {\bibfnamefont {D.}~\bibnamefont
  {Abanin}}, \bibinfo {author} {\bibfnamefont {W.~De}\ \bibnamefont {Roeck}},
  \bibinfo {author} {\bibfnamefont {W.~W.}\ \bibnamefont {Ho}}, \ and\ \bibinfo
  {author} {\bibfnamefont {F.}~\bibnamefont {Huveneers}},\ }\bibfield  {title}
  {\enquote {\bibinfo {title} {A rigorous theory of many-body prethermalization
  for periodically driven and closed quantum systems},}\ }\href {\doibase
  10.1007/s00220-017-2930-x} {\bibfield  {journal} {\bibinfo  {journal}
  {Commun. Math. Phys.}\ }\textbf {\bibinfo {volume} {354}},\ \bibinfo {pages}
  {809--827} (\bibinfo {year} {2017})}\BibitemShut {NoStop}%
\bibitem [{\citenamefont {Benatti}\ \emph {et~al.}(2018)\citenamefont
  {Benatti}, \citenamefont {Carollo}, \citenamefont {Floreanini},\ and\
  \citenamefont {Narnhofer}}]{Benatti_2018}%
  \BibitemOpen
  \bibfield  {author} {\bibinfo {author} {\bibfnamefont {F.}~\bibnamefont
  {Benatti}}, \bibinfo {author} {\bibfnamefont {F.}~\bibnamefont {Carollo}},
  \bibinfo {author} {\bibfnamefont {R.}~\bibnamefont {Floreanini}}, \ and\
  \bibinfo {author} {\bibfnamefont {H.}~\bibnamefont {Narnhofer}},\ }\bibfield
  {title} {\enquote {\bibinfo {title} {Quantum spin chain dissipative
  mean-field dynamics},}\ }\href {\doibase 10.1088/1751-8121/aacbdb} {\bibfield
   {journal} {\bibinfo  {journal} {J. Phys. A: Math. Theor.}\ }\textbf
  {\bibinfo {volume} {51}},\ \bibinfo {pages} {325001} (\bibinfo {year}
  {2018})}\BibitemShut {NoStop}%
\bibitem [{\citenamefont {Kshetrimayum}\ \emph {et~al.}(2017)\citenamefont
  {Kshetrimayum}, \citenamefont {Weimer},\ and\ \citenamefont
  {Or{\'u}s}}]{Kshetrimayum:2017aa}%
  \BibitemOpen
  \bibfield  {author} {\bibinfo {author} {\bibfnamefont {A.}~\bibnamefont
  {Kshetrimayum}}, \bibinfo {author} {\bibfnamefont {H.}~\bibnamefont
  {Weimer}}, \ and\ \bibinfo {author} {\bibfnamefont {R.}~\bibnamefont
  {Or{\'u}s}},\ }\bibfield  {title} {\enquote {\bibinfo {title} {A simple
  tensor network algorithm for two-dimensional steady states},}\ }\href
  {\doibase 10.1038/s41467-017-01511-6} {\bibfield  {journal} {\bibinfo
  {journal} {Nat. Commun.}\ }\textbf {\bibinfo {volume} {8}},\ \bibinfo {pages}
  {1291} (\bibinfo {year} {2017})}\BibitemShut {NoStop}%
\bibitem [{\citenamefont {Carollo}\ \emph {et~al.}(2019)\citenamefont
  {Carollo}, \citenamefont {Gillman}, \citenamefont {Weimer},\ and\
  \citenamefont {Lesanovsky}}]{PhysRevLett.123.100604}%
  \BibitemOpen
  \bibfield  {author} {\bibinfo {author} {\bibfnamefont {F.}~\bibnamefont
  {Carollo}}, \bibinfo {author} {\bibfnamefont {E.}~\bibnamefont {Gillman}},
  \bibinfo {author} {\bibfnamefont {H.}~\bibnamefont {Weimer}}, \ and\ \bibinfo
  {author} {\bibfnamefont {I.}~\bibnamefont {Lesanovsky}},\ }\bibfield  {title}
  {\enquote {\bibinfo {title} {Critical behavior of the quantum contact process
  in one dimension},}\ }\href {\doibase 10.1103/PhysRevLett.123.100604}
  {\bibfield  {journal} {\bibinfo  {journal} {Phys. Rev. Lett.}\ }\textbf
  {\bibinfo {volume} {123}},\ \bibinfo {pages} {100604} (\bibinfo {year}
  {2019})}\BibitemShut {NoStop}%
\bibitem [{\citenamefont {Bocchieri}\ and\ \citenamefont
  {Loinger}(1957)}]{PhysRev.107.337}%
  \BibitemOpen
  \bibfield  {author} {\bibinfo {author} {\bibfnamefont {P.}~\bibnamefont
  {Bocchieri}}\ and\ \bibinfo {author} {\bibfnamefont {A.}~\bibnamefont
  {Loinger}},\ }\bibfield  {title} {\enquote {\bibinfo {title} {Quantum
  recurrence theorem},}\ }\href {\doibase 10.1103/PhysRev.107.337} {\bibfield
  {journal} {\bibinfo  {journal} {Phys. Rev.}\ }\textbf {\bibinfo {volume}
  {107}},\ \bibinfo {pages} {337--338} (\bibinfo {year} {1957})}\BibitemShut
  {NoStop}%
\bibitem [{\citenamefont {Deutsch}(1991)}]{PhysRevA.43.2046}%
  \BibitemOpen
  \bibfield  {author} {\bibinfo {author} {\bibfnamefont {J.~M.}\ \bibnamefont
  {Deutsch}},\ }\bibfield  {title} {\enquote {\bibinfo {title} {Quantum
  statistical mechanics in a closed system},}\ }\href {\doibase
  10.1103/PhysRevA.43.2046} {\bibfield  {journal} {\bibinfo  {journal} {Phys.
  Rev. A}\ }\textbf {\bibinfo {volume} {43}},\ \bibinfo {pages} {2046--2049}
  (\bibinfo {year} {1991})}\BibitemShut {NoStop}%
\bibitem [{\citenamefont {Srednicki}(1994)}]{PhysRevE.50.888}%
  \BibitemOpen
  \bibfield  {author} {\bibinfo {author} {\bibfnamefont {M.}~\bibnamefont
  {Srednicki}},\ }\bibfield  {title} {\enquote {\bibinfo {title} {Chaos and
  quantum thermalization},}\ }\href {\doibase 10.1103/PhysRevE.50.888}
  {\bibfield  {journal} {\bibinfo  {journal} {Phys. Rev. E}\ }\textbf {\bibinfo
  {volume} {50}},\ \bibinfo {pages} {888--901} (\bibinfo {year}
  {1994})}\BibitemShut {NoStop}%
\bibitem [{\citenamefont {Polkovnikov}\ \emph {et~al.}(2011)\citenamefont
  {Polkovnikov}, \citenamefont {Sengupta}, \citenamefont {Silva},\ and\
  \citenamefont {Vengalattore}}]{RevModPhys.83.863}%
  \BibitemOpen
  \bibfield  {author} {\bibinfo {author} {\bibfnamefont {A.}~\bibnamefont
  {Polkovnikov}}, \bibinfo {author} {\bibfnamefont {K.}~\bibnamefont
  {Sengupta}}, \bibinfo {author} {\bibfnamefont {A.}~\bibnamefont {Silva}}, \
  and\ \bibinfo {author} {\bibfnamefont {M.}~\bibnamefont {Vengalattore}},\
  }\bibfield  {title} {\enquote {\bibinfo {title} {Colloquium: Nonequilibrium
  dynamics of closed interacting quantum systems},}\ }\href {\doibase
  10.1103/RevModPhys.83.863} {\bibfield  {journal} {\bibinfo  {journal} {Rev.
  Mod. Phys.}\ }\textbf {\bibinfo {volume} {83}},\ \bibinfo {pages} {863--883}
  (\bibinfo {year} {2011})}\BibitemShut {NoStop}%
\bibitem [{\citenamefont {D'Alessio}\ \emph {et~al.}(2016)\citenamefont
  {D'Alessio}, \citenamefont {Kafri}, \citenamefont {Polkovnikov},\ and\
  \citenamefont {Rigol}}]{dalessio2016}%
  \BibitemOpen
  \bibfield  {author} {\bibinfo {author} {\bibfnamefont {L.}~\bibnamefont
  {D'Alessio}}, \bibinfo {author} {\bibfnamefont {Y.}~\bibnamefont {Kafri}},
  \bibinfo {author} {\bibfnamefont {A.}~\bibnamefont {Polkovnikov}}, \ and\
  \bibinfo {author} {\bibfnamefont {M.~A.}\ \bibnamefont {Rigol}},\ }\bibfield
  {title} {\enquote {\bibinfo {title} {From quantum chaos and eigenstate
  thermalization to statistical mechanics and thermodynamics},}\ }\href
  {\doibase 10.1080/00018732.2016.1198134} {\bibfield  {journal} {\bibinfo
  {journal} {Adv. Phys.}\ }\textbf {\bibinfo {volume} {65}},\ \bibinfo {pages}
  {239--362} (\bibinfo {year} {2016})}\BibitemShut {NoStop}%
\bibitem [{\citenamefont {Zwolak}\ and\ \citenamefont
  {Vidal}(2004)}]{Zwolak2004}%
  \BibitemOpen
  \bibfield  {author} {\bibinfo {author} {\bibfnamefont {M.}~\bibnamefont
  {Zwolak}}\ and\ \bibinfo {author} {\bibfnamefont {G.}~\bibnamefont {Vidal}},\
  }\bibfield  {title} {\enquote {\bibinfo {title} {Mixed-state dynamics in
  one-dimensional quantum lattice systems: A time-dependent superoperator
  renormalization algorithm},}\ }\href {\doibase 10.1103/PhysRevLett.93.207205}
  {\bibfield  {journal} {\bibinfo  {journal} {Phys. Rev. Lett.}\ }\textbf
  {\bibinfo {volume} {93}},\ \bibinfo {pages} {207205} (\bibinfo {year}
  {2004})}\BibitemShut {NoStop}%
\bibitem [{\citenamefont {Verstraete}\ \emph {et~al.}(2004)\citenamefont
  {Verstraete}, \citenamefont {Garc\'{\i}a-Ripoll},\ and\ \citenamefont
  {Cirac}}]{Verstraete2004}%
  \BibitemOpen
  \bibfield  {author} {\bibinfo {author} {\bibfnamefont {F.}~\bibnamefont
  {Verstraete}}, \bibinfo {author} {\bibfnamefont {J.~J.}\ \bibnamefont
  {Garc\'{\i}a-Ripoll}}, \ and\ \bibinfo {author} {\bibfnamefont {J.~I.}\
  \bibnamefont {Cirac}},\ }\bibfield  {title} {\enquote {\bibinfo {title}
  {Matrix product density operators: Simulation of finite-temperature and
  dissipative systems},}\ }\href {\doibase 10.1103/PhysRevLett.93.207204}
  {\bibfield  {journal} {\bibinfo  {journal} {Phys. Rev. Lett.}\ }\textbf
  {\bibinfo {volume} {93}},\ \bibinfo {pages} {207204} (\bibinfo {year}
  {2004})}\BibitemShut {NoStop}%
\end{thebibliography}%

\section*{Methods}
\subsection*{Mean-field regime for the Rydberg dissipative dynamics}
\label{app2}
To study the dissipative system in the mean-field approximation, i.e. neglecting correlations between atoms, it is sufficient to enforce a tensor product structure of the quantum state. Given that the initial state, assumed to be the vacuum state, is translation invariant, and the same is true for the dynamical generator, all atoms are described by the same reduced state and thus the only relevant degrees of freedom of the system (given that correlations are neglected) are $s_x=\langle \sigma_x\rangle$, $s_y=\langle \sigma_y\rangle$, and  the density of excitations $s_n=\langle n\rangle$. The dynamics of these degrees of freedom is generated by the non-linear differential equations  \cite{Benatti_2018}
\begin{equation}
\begin{split}
\dot{s}_x&=\mu s_{y}-\frac{2}{v}s_ns_y-\frac{\gamma}{2}s_x\, ,\\
\dot{s}_y&=-2\Omega(2s_n-1)-\mu s_x+\frac{2}{v}s_ns_x-\frac{\gamma}{2}s_y\,,\\
\dot{s}_n&=\Omega s_y-\gamma s_n\, .
\end{split}
\end{equation}
These equations describe the evolution for a given choice of the Hamiltonian and dissipative parameter. To perform the engine cycle, we have to periodically change the parameters, as explained in the main text, providing as initial state for each transformation the final state of the previous one. As the first initial state, we chose the state with all atoms in their (zero-excitation) ground state $|g\rangle$.

The Rydberg interaction energy is given, in the mean-field approximation and in thermodynamic limit of infinitely many atoms, by 
$$
\lim_{N_{\rm at}\to\infty}\frac{1}{N_{\rm at}}\langle H^v_{\rm LG}\rangle = \frac{1}{v}s_n^2\, .
$$\\

\subsection*{Results for an infinite 1D dissipative Rydberg chain}
\label{app4}
To obtain results for an infinite 1D Rydberg system we exploit ITEBD algorithms \cite{Vidal2007,paeckel2019timeevolution} which permit, for the  dynamics in Eq.~\eqref{Lindblad}, an efficient approximation of the translation-invariant stationary many-body density matrix by running real-time dynamics as done, e.g., in Refs.~\cite{Kshetrimayum:2017aa,PhysRevLett.123.100604}. In particular, given a pair $\mu,v$, we can use such an algorithm to estimate the stationary value of the corresponding density-density nearest-neighbour interactions $\langle n^{(k)} n^{(k+1)}\rangle$. Repeating this operation for a sufficiently fine grid, we obtain an estimate for $\langle n^{(k)} n^{(k+1)}\rangle$ as a function of $\mu$ and $v$. Since for 1D systems, the Lindblad equation \eqref{Lindblad} is expected to have a unique steady-state for any pair of $\mu,v$, the stationary value of $\langle n^{(k)} n^{(k+1)}\rangle$ does not depend on the initial conditions. After both quenches in $\mu$ the system relaxes to stationarity and thus the values $\langle n^{(k)} n^{(k+1)}\rangle_{A/C}$ depend only on the actual values of $\mu,v$. These values, which can be found from the table function constructed for $\langle n^{(k)} n^{(k+1)}\rangle$, can be used to infer exact properties of the stationary cycle. Summarizing, the procedure provides exact results apart from the approximation of the stationary state of the Lindblad equation done via matrix product states and the calculation of $\langle n^{(k)} n^{(k+1)}\rangle_{A/C}$ via interpolation from the table function constructed through a sufficient number of $\mu,v$ points. We notice that, contrary to the discussion  in Ref.~\cite{PhysRevLett.123.100604}, here the approximation of the density-density interactions converges quite fast; this confirms the expectation that no phase transition occurs in the 1D chain. Overall, this procedure allows for an efficient investigation of the phase diagram in Fig.~\ref{Fig4}(b).

\onecolumngrid
\newpage

\renewcommand\thesection{S\arabic{section}}
\renewcommand\theequation{S\arabic{equation}}
\renewcommand\thefigure{S\arabic{figure}}
\setcounter{equation}{0}
\setcounter{figure}{0}

\begin{center}
{\Large SUPPLEMENTARY INFORMATION}
\end{center}
\begin{center}
\vspace{0.8cm}
{\Large A non-equilibrium quantum many-body Rydberg atom engine}
\end{center}
\begin{center}
Federico Carollo,$^{1}$ Filippo M. Gambetta,$^{1}$ Kay Brandner,$^{2}$ Juan P. Garrahan$^{1}$ and Igor Lesanovsky$^{1,3}$
\end{center}
\begin{center}
$^1${\em School of Physics and Astronomy and}\\
{\em Centre for the Mathematics and Theoretical Physics of Quantum Non-Equilibrium Systems,}\\
{\em  University of Nottingham, Nottingham, NG7 2RD, UK}\\
$^2${\em Department of Physics, Keio University, 3-14-1 Hiyoshi, Yokohama, 223-8522, Japan}\\
$^3${\em Institut f\"ur Theoretische Physik, Universit\"at T\"ubingen,}\\
{\em Auf der Morgenstelle 14, 72076 T\"ubingen, Germany}
\end{center}

\section{Exchanged work in a sudden quench transformation}
\label{app0}
In this section, we derive the expression for the work associated with sudden quenches of the specific volume $v$ from the general equation \eqref{ft-work} in the main text. For a single transformation involving a volume change, we have
$$
w:=\int dt \, f_t^{v_t}\dot{v}_t= \frac{1}{N_{\rm at}}\int_0^{t'}\langle \frac{\partial }{\partial v} H^v_{\rm LG}\Big|_{v=v_t}\rangle \dot{v}_t dt\, . 
$$
For a sudden quench, we have that 
$$
v(0)=v_-\, , \qquad v(0^+)=v_+\, ,
$$
and $v(t)=v_+$, for $t>0^+$. Thus, irrespective of the duration of the transformation, the integral reduces to
$$
w=-\int_0^{0^{+}} dt \frac{dv(t)}{dt}\left(\frac{1}{v^2(t)}\right)\frac{1}{N_{\rm at}}\langle \sum_{k=1}^{N_{\rm at}-1}n^{(k)}n^{(k+1)}\rangle (t)\, .
$$
Changing the integration variable to $v$ and using that from $0$ to $0^+$ there is no actual evolution, we obtain  
$$
\langle \sum_{k=1}^{N_{\rm at}-1}n^{(k)}n^{(k+1)}\rangle (t)=\langle \sum_{k=1}^{N_{\rm at}-1}n^{(k)}n^{(k+1)}\rangle (0)=\langle \sum_{k=1}^{N_{\rm at}-1}n^{(k)}n^{(k+1)}\rangle\, ,
$$
and thus  
$$
w=-\frac{1}{N_{\rm at}}\langle \sum_{k=1}^{N_{\rm at}-1}n^{(k)}n^{(k+1)}\rangle \int_{v(0)}^{v(0^+)} dv\frac{1}{v^2}\, ,
$$
and finally 
\begin{equation}
w=\frac{1}{N_{\rm at}}\langle \sum_{k=1}^{N_{\rm at}-1}n^{(k)}n^{(k+1)}\rangle \left(\frac{1}{v_+}-\frac{1}{v_-}\right)\, .
\label{Work-supp}
\end{equation}\\

\section{Periodic driving  and dissipation}

\begin{figure}[h]
\centering
\includegraphics[scale=0.35]{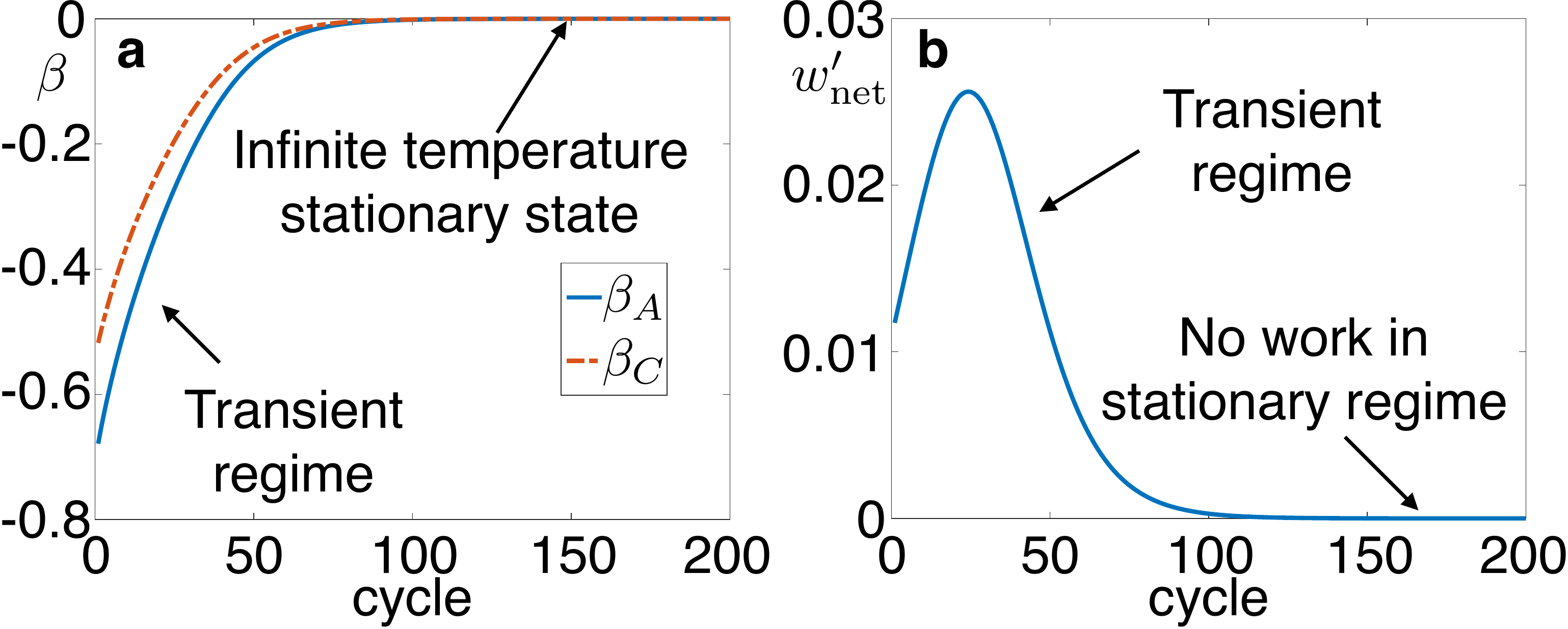}
\caption{{\bf Closed infinite 1D Rydberg chain. } Fate of the Rydberg engine under fully coherent, i.e.~dissipationless, driving. The parameters for both plots are $\Omega=1$, $v_{\rm min}=0.7$, $\mu_{\rm min}=4$, and $\Delta v=\Delta \mu =1$. {\bf a} Effective inverse temperatures $\beta_A$, $\beta_C$ describing the thermalised many-body state at the parameter configurations  $A$ and $C$ --appearing in Fig.~\ref{Fig2}(a) of the main text-- as a function of the number of driving cycles performed. 
After a transient regime, the system heats up to infinite temperature. {\bf b} Consequently, the stationary cycle becomes trivial and work can be extracted only within a transient regime. The limit state is equivalent to an  infinite temperature state (proportional to the identity matrix) and  the net work vanishes, $w_{\rm net}=0$. For different parameters (not displayed) it might not even be possible to extract work during transient regimes.  }
\label{Fig3}
\end{figure}

In the absence of dissipation the evolution of any initial state $|\psi_0\rangle$  is described by the Schr\"odinger equation. Hence, for any finite system, one will eventually observe Poincar\'e recurrence \cite{PhysRev.107.337}. The emergence of stationary regimes, desirable to stabilise the thermodynamic cycle, will therefore only be achievable in the limit of infinitely many particles and infinitely long times \cite{PhysRevA.43.2046,PhysRevE.50.888,RevModPhys.83.863,dalessio2016}. However, it is generically expected that periodically-driven unitary quantum systems will heat up to infinite temperature \cite{PhysRevX.4.041048,PhysRevE.90.012110,PONTE2015196} making it impossible  to sustain a non-trivial engine cycle ($w_{\rm net}'=0$). Indeed, exploiting the eigenstate thermalization hypothesis (ETH) \cite{PhysRevA.43.2046,PhysRevE.50.888,RevModPhys.83.863,dalessio2016}, we verify this behaviour for our system considering infinitely many particles and infinitely long relaxation times  (see section \ref{app1}). In the best case scenario, we find  that work can be produced, for a periodically-driven isolated  Rydberg system, in a transient regime. At stationarity, the effective inverse temperatures $\beta_A$ and $\beta_C$, shown in Fig.~\ref{Fig3}(a), which characterize the thermalised many-body state, tend to zero and  no net work is produced [see Fig.~\ref{Fig3}(b)].

Extracting net work thus requires a dynamical mechanism that prevents this thermal catastrophe. Such a mechanism is naturally given, for our Rydberg engine, by radiative decay. This process leads to  non-equilibrium steady-states which do not feature an infinite temperature, as demonstrated by the fact that $\mathcal{L}[{\bf 1}]\neq 0$, when $\gamma >0$. The system state is thus expected to converge to a non-trivial periodic cycle. 

\section{ The thermodynamic cycle in closed systems at large scales}
\label{app1}

In order to show the heating  to infinite temperature for the dissipationless system,  we devise an ad-hoc algorithm, that is based on the eigenstate thermalization hypothesis (ETH) and exploits infinite time-evolving block-decimation (ITEBD) algorithms. 

The evolution of an initial state $|\psi_0\rangle$ under the Hamiltonian $H$ is given by $|\psi_t\rangle=e^{-i Ht}|\psi_0\rangle$. Since the Hamiltonian is non-integrable the system is, in the thermodynamic limit and for very long times,  generically expected to thermalise \cite{PhysRevA.43.2046,PhysRevE.50.888,RevModPhys.83.863,dalessio2016} to a Gibbs state with respect to an appropriate inverse temperature $\beta$. Hence, for an infinite Rydberg chain, one should have
\begin{equation}
\lim_{t\to \infty}\langle\psi_t|O|\psi_t\rangle=\Tr\left(\rho_\beta \,O\right)\, ,\qquad \rho_\beta\propto e^{-\beta\, H}
\label{ETH}
\end{equation}
for any  local observables $O$, and for $|\psi_0\rangle$ being sufficiently localised in energy. The inverse temperature $\beta$ is constrained by energy conservation
\begin{equation}
\langle \psi_0|H|\psi_0\rangle=\Tr\left(\rho_\beta \,H\right)\, .
\label{Energy-constraint}
\end{equation}
This mathematical property entails thermalisation of the quantum state on long time-scales. The simple energy constraint above holds for a sudden quench of any Hamiltonian parameter, which is the case under investigation in our paper. 

To explain how our algorithm works, it is sufficient to consider a single transformation. Each transformation begins with a sudden change of the parameters in the Hamiltonian; if we are able to compute the energy (with respect to $H$) in the system just after the quench we can obtain the value of $\beta$ for the final state of the transformation (in an infinite relaxation time) exploiting ETH and the energetic constraints given in Eq.~\eqref{Energy-constraint}. 
Moreover, the knowledge of  $\beta$ identifies the thermal state $\rho_\beta$ and thus makes it possible to compute  the expectation value of lattice gas Hamiltonian, $H_{\rm LG}^v$, which is necessary to determine work exchanges in the cycle, as well as  expectation values of local observables; the latter are required to compute the energy injected in the following cycle. Note that, even if effective Gibbs states are involved in the computation of expectations, the system state is, by construction, pure at any time during the cycle. Thus, assuming that at every point in the thermodynamic cycles the ETH is valid  (this is an assumption of the algorithm), one can  repeat the above steps for all transformations. The role of tensor networks, and precisely of iMPS \cite{Vidal2007} for 1D systems, is to estimate expectation values, i.e.~, the average value of local observables in the thermal state $\rho_\beta$ \cite{Zwolak2004,Verstraete2004}. By means of this algorithm, we are able to make estimates for the cycle in the limit of infinite relaxation after the quenches in $v$ and $\mu$ (which is enabled by the ETH assumption) and in the limit of infinitely many particles (enabled by tensor networks). \\

\section{Supplementary results for finite Rydberg chain}
\label{app5} 

\begin{figure}[h]
\centering
\includegraphics[scale=0.35]{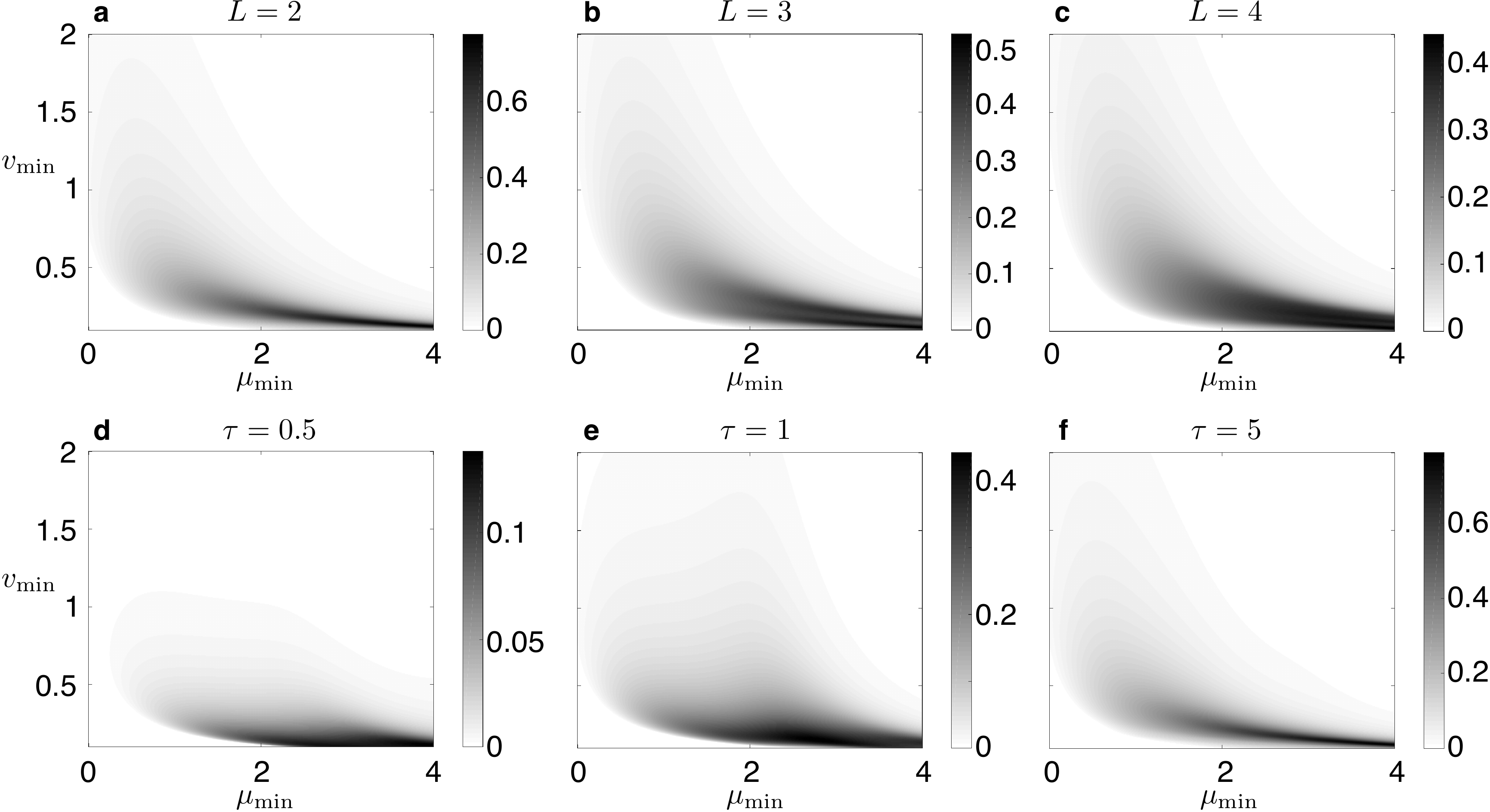}
\caption{{\bf Results for finite systems and finite relaxation times. } Density plot of the net work delivered by the cycle with same parameters  as in Fig.~\ref{Fig4} of the main text. Here we consider a finite system with open boundary conditions (panels {\bf a}-{\bf c}) with $L=2,3,4$ and finite relaxation times $\tau=0.5,1,5$ (panels {\bf d}-{\bf f}) for a system with $L=2$. The  net work is obtained by analysing the stable stationary cycle. All  plots show that it is possible to extract work also in few-body Rydberg systems and with finite power. The bottom left region of each plot, as  in Fig.~\ref{Fig4} of the main text, is a region of negative net work which has been set to zero. }
\label{Fig5}
\end{figure}

Here we report supplementary results on the Rydberg engine cycle of the main text. Specifically, we show that it is possible to extract work from the Rydberg atomic array even for  few particles and for  finite relaxation times after the quenches on the parameter $\mu$. This conclusion is clearly supported  by numerical results obtained from exact diagonalisation, which are displayed in Fig.~\ref{Fig5}. To account for finite-size effects, we compute the work during a transformation using Eq.~\eqref{Work-supp}, where  we divide the expectation value for the lattice gas Hamiltonian, $H^v_{\rm LG}$, by the number of bonds $N_{\rm at}-1$ instead of the number of atoms. In the large size limit ($N_{\rm at}\to\infty$) this difference is not relevant.

\end{document}